\documentclass[12pt]{article}
\usepackage[margin=1.0in]{geometry}
\usepackage[utf8]{inputenc}
\usepackage{amsthm,amsmath,physics,mathtools,amssymb}

\usepackage{charter}

\usepackage{CJKutf8}

\usepackage{enumitem}

\usepackage{hyperref}
\hypersetup{
    colorlinks=true,
    linkcolor=blue,
    filecolor=magenta,      
    urlcolor=cyan,
}
\usepackage[capitalise]{cleveref}

\theoremstyle{definition}
\newtheorem{theorem}{Theorem} 
\newtheorem{proposition}[theorem]{Proposition} 

\newtheorem{example}{Example}

\usepackage{graphicx}
\graphicspath{ {images/} }

\usepackage{tensor}

\DeclarePairedDelimiterX{\inp}[2]{\langle}{\rangle}{#1, #2}

\usepackage{xparse}

\NewDocumentCommand\LH{mo}{%
  \IfNoValueTF{#2}
   {\mathcal{L}(\mathcal{H}^{#1})}
   {\mathcal{L}(\mathcal{H}^{#1},\mathcal{H}^{#2})}%
}

\newcommand\id{\leavevmode\hbox{\small1\kern-3.3pt\normalsize1}}

\allowdisplaybreaks

\usepackage[T1]{fontenc}

\usepackage{bbold}

\usepackage{authblk}

\usepackage[title]{appendix}

\usepackage{mciteplus}

\usepackage{upgreek}

\usepackage{epigraph}

\usepackage{mdframed,lipsum}
\mdfdefinestyle{MyFrame}{%
    linecolor=black,
    outerlinewidth=2pt,
    innertopmargin=4pt,
    innerbottommargin=4pt,
    innerrightmargin=4pt,
    innerleftmargin=4pt,
        leftmargin = 4pt,
        rightmargin = 4pt
        }

\title{Modes of experience in a superposed world}

\author{Ding Jia (贾丁)\thanks{djia@perimeterinstitute.ca}}
\affil{Perimeter Institute for Theoretical Physics, Waterloo, Ontario, N2L 2Y5, Canada}
\affil{Department of Physics and Astronomy, University of Waterloo, Waterloo, Ontario, N2L 3G1, Canada}
\date{}

\begin{document}

\begin{CJK*}{UTF8}{gbsn}
\maketitle
\end{CJK*}

\begin{abstract}
A central open problem of quantum physics is to reconcile theory with experience. In this work I present a framework for studying distinct modes of experience in a superposed world. A modes of experience is characterized by how the world, experiences, and options relate to each other by the perceptions, decisions, and actions, as well as by probabilistic rules encoding probabilistic or deterministic correlations among first person experiences. In a toy model, the life expectancies of beings in different candidate modes of experience are compared. It is found that the quantum mode without macroscopic superposition outlives that with macroscopic superposition and that with real amplitudes. These highlight the prospect to explain a mode of experience by its evolutionary advantages.
\end{abstract}

\section{Introduction}

\setlength{\epigraphwidth}{0.9
\textwidth}
\epigraph{He once greeted me with the question: ``Why do people say that it was natural to think that the sun went round the earth rather than that the earth turned on its axis?'' I replied: ``I suppose, because it looked as if the sun went round the earth.'' ``Well,'' he asked, ``what would it have looked like if it had looked as if the earth turned on its axis?''}{G. E. M. Anscombe telling a story about L. Wittgenstein \cite{Anscombe1963AnTractatus}}

``Why do people say that quantum theory is weird?'' ``I suppose, because it describes a superposed world, but we do not experience the superpositions.'' ``Well, what would we experience if the world was a superposed world?''

Indeed, what would we experience if the world was a superposed world? Would we experience superpositions? Would we not experience superpositions? What do we mean by ``we''? Does it include octopuses? Does it include pigeons? Does it include aliens? Does it include quantum computers? Do the answers differ for these different kinds?

Dogs have a wider hearing range than humans. There are sounds that dogs hear but humans do not. This illustrates two important points. First, not all that exist are experienced.\footnote{One might also ask whether any existence is experienced at all \cite{Hoffman2015ThePerception}.} Second, what is experienced depends on species.

In an attempt to tell what would be experienced if the world was a superposed world, a physicist immediately encounter the \textbf{experience question}:
\begin{mdframed}[style=MyFrame,nobreak=true,align=center,userdefinedwidth=23em]
How to relate the theory of physics to experience?
\end{mdframed}
Arguably, to relate physical existence to experience is already an open problem in classical physics (e.g., can one tell from the entire physical configuration of a classical world what a person perceives from a Necker cube?). Certainly in quantum physics the problem is still open. In the words of Page \cite{PageClassicalityConsciousness}:
\begin{quote}
[...] the quantum state and dynamics do not by themselves logically imply what (if any) conscious perceptions occur, or how much.
\end{quote} 
Another challenge is the \textbf{multiplicity question}:
\begin{mdframed}[style=MyFrame,nobreak=true,align=center,userdefinedwidth=23em]
How do experience vary from species to species? From individuals to individuals? From different stages of the same individual?
\end{mdframed}
If it is true that all lives on Earth evolved from rudimentary organisms in the sea, then it is unlikely that the current mode of experience of human beings apply to all experiential beings that existed, exist, and will exist. In the cosmos, if is even less likely that all experiential beings experience a superposed world in the same mode. Therefore a general answer to the question ``what would be experienced if the world was a superposed world'' must account for the multiplicities of modes of experience.

Obviously the question is difficult. In this work, I focus on the sub-question of whether evolutionary considerations could shed light on explaining modes of experience in a superposed world. In a toy model, I compare the life expectancy of beings without experiencing macroscopic superposition, to the life expectancy of beings in some other hypothetical modes of experience. It is found that the first mode of experience outlives the others. This demonstrates the potential to apply evolutionary reasoning to make progress on questions about experience in superposed world. 

To set up the study, I needed to discuss what it could mean to talk about a ``superposed world''. Then adapting the language of psychologists/cognitive scientists, I set up a framework to formulate (candidate) modes of experiences in a superposed world in a language closer to that of physicists. This General Experience Theories (GET) language (reminiscent (see \Cref{sec:fgpt}) of the General Probabilistic Theories (GPT) \cite{Popescu1994QuantumAxiom, HardyQuantumAxioms, barrett_information_2007} ) may serve as a starting point in further studies of experience in a superposed world.

As some examples of additional questions to study in this GET language, how would beings with experiences of macroscopic superposition communicate among themselves? How would they appear to and communicate with other beings that do not perceive macroscopic superpositions? What kind of traces would beings with alternative modes of experience leave if they existed in the past but did not survive? Can some probability rules of general probabilistic theories originally postulated for hypothetical universes actually apply to alternative modes of experience in our universe? Can some beings be said to have Free Will, and if so in which sense? The list goes on...

The potential relevance of evolution in the modes of experience reveals a distinct perspective on the ontological and interpretation issues of quantum physics (\Cref{sec:taqt}). Once it is recognized that the current human mode of experience may not apply to all experiences in a superposed world, our understandings on the ontology of a superposed world, on the nature of probabilities in quantum theory, and on the realm of applicability of ordinary quantum probabilities rules need to be questioned and updated.

The paper is organized as follows. In \Cref{eq:rw}, I list some related previous works. In \Cref{sec:ennee}, I explain why experiences should not be expected to encompass physical existence, which motivates a treatment that separates the set of experiences and the set of physical world configurations. In \Cref{sec:sw}, I make precise the notion of ``superposed world'' in the current discussion as an objective description of the world in terms of the totality of path integral configurations in superposition, but without the amplitudes. In \Cref{sec:moe}, I introduce a language to formalize candidate modes of experience in a superposed world and offer several examples. In \Cref{sec:ec}, I study in three distinct candidate modes of experience in a toy universe of 1D scalar field theory. It is shown that beings in the ordinary quantum mode have longer life expectancy than beings in the mode of macroscopic superpositions and beings in the mode of real amplitudes. In \Cref{sec:d}, I discuss some broader topics in the present setting: ontology; the nature of probability; theories of everything; General Probabilistic Theories and General Experience Theories. Based on these discussions, I conclude with some tentative ideas towards an interpretation of quantum theory.

\subsection{Related works}\label{eq:rw}

Here I list some works related to the topic of experience in a superposed world from previous literature. This is by no means a complete list, as I only mention the works that I referred to for the study.

The experience question is certainly an old one. In quantum physics it is intimately related to the measurement problem. Kent's ``Night thoughts of a quantum physicist'' \cite{Kent2000NightPhysicist} offers the most convincing argument for studying experience/consciousness to make real progress in quantum physics I have seen. The book \cite{2022ConsciousnessMechanics} edited by Shan Gao contains several very interesting recent discussions about experience in quantum physics. For overall discussions on the relevance of understanding experience to understand quantum physics, see for instance Shan Gao's and Jenann Ismael's entries, and Peter Lewis' entry for a discussion on positive and negative ways of mentioning ``experience'' in quantum theory.

Don Page's ``Sensible Quantum Mechanics'' \cite{PageSensibleProbabilistic, Page1995SensibleMind, Page2003MindlessConsciousness} puts conscious experience at center stage in quantum physics. Quentin's, Lockwood's, and Loewer's entries in the same book of \cite{Page2003MindlessConsciousness}, as well as Squires' \cite{Squires1990ConsciousWorld, Squires1993QuantumWorld} offer interesting discussions on the related topics. 

Ted Chiang's science fiction \textit{Story of Your Life} \cite{Chiang1998StoryLife} and the associated film \textit{Arrival} \cite{Villeneuve2016Arrival} depict an alien species with a very different mode of experience. While for human beings experiences unfold sequentially in time, for the aliens experiences occur ``all at once''. Viewed from a human being's perspective, the aliens know of what happens later at the present, as if they have ``memories of the future''.

Although this story is fictional, the overall point that another species can have a quite distinct mode of experience has its factual basis. As discussed in Peter Godfrey-Smith's intriguing book \textit{Other Minds} \cite{Godfrey-Smith2016OtherConsciousness}, Darwinian evolution on Earth itself have already created animals with quite distinct perceptive systems. For example, not all animals possess a centralized neural system concentrated at the brain as humans do. Octopuses actually have most of their neurons in the arms. The arms even enjoy a certain degree of independence, and can autonomously touch, taste and move without inputs from the brain. Even some mammals do not integrate their experience nearly as much as human beings do, and exhibit fissures in information processing associated with the separation between the two halves of the brain. For example, pigeons can learn and achieve some simple tasks such as discriminating two shapes with one eye covered. Yet when pushed to perform the same task with the covered eye they tend to fail, as if the skill learned by half of the brain does not carry to the other half.

That different beings may experience a superposed world differently is alluded to by Hartle. Referring to information gathering and utilizing systems (IGUSes) which contain different experiential beings as special cases, Hartle raises the question \cite{Hartle2011TheUniverse}:
\begin{quote}
Could the quasiclassical realms of this universe contain quasiclassically described IGUSes elsewhere whose senses register variables substantially different from the ones we use, even non-quasiclassical ones? 
To answer it would be necessary to calculate the probabilities of alternative evolutionary histories of such quasiclassically
described IGUSes. It is well beyond our power at present to even formulate such
a calculation precisely much less carry it out. If we ever encounter extra-terrestrial
IGUSes this question may be settled experimentally.
\end{quote}
Recently, Brodutch \textit{et al.} \cite{BrodutchDoSheep} discuss the possibility of observers with quantum memories. It is speculated that sufficiently advanced fault-tolerant quantum computers with peripheral quantum sensors can constitute such observers who perform generalized measurements that access quantum states without reduction.

Some attempts to derive the Born rule are critically reviewed in Landsman \cite{Landsman2009TheInterpretation} and Vaidman \cite{Vaidman2020DerivationsRule}, and reasons to be dissatisfied with these derivations are pointed out. These are related to the current study, because if indeed distinct modes of experience may be realized in Nature, we will never be able to derive the human mode of experience directly from the universal laws of physics, just like we will never derive that a living being must have two legs directly from the universal laws of biology. Because the universal laws also apply to other distinct modes of experience, the mode of humans could only be explained by identifying additional inputs contingent on our species.

Hoffman \textit{et al.}'s works of the ``interface theory of perception'' \cite{Hoffman2014ObjectsConsciousness, Hoffman2015ThePerception, Hoffman2015ProbingCommentaries} advocates the view that perception need not reflect existence. Related to this topic \cite{Fields2017EigenformsEncoding}, Chris Fields has written very interesting works on the interaction of systems including experiential ones applicable to quantum physics \cite{Fields2013ATheory, Fields2016BuildingWorld}. Markus M{\"u}ller's work on a first-person approach to laws of physics \cite{Muller2020LawTheory} has also been very inspiring for my work. Ideas of non-obvious relationships between existence and experience may strike the uninitiated as surprising or exotic. Yet to some initiated in thinking about the experience question, the same ideas are natural and even mundane \cite{Heyer2002PerceptionWorld, Mausfeld2015NotionsPerception}. See \cite{Dehaene2014ConsciousnessThoughts, Frith2007MakingWorld, Lindsay2021ModelsMind} for some delightful introductions to progresses in recent decades on the subject of human perception.

One point stressed by Hoffman \textit{et al.} is that ``perception is about having kids, not seeing truth'' \cite{Hoffman2015ThePerception}.
The relevance of evolutionary fitness to quantum foundations has been noted before. Zurek remarks that \cite{Zurek2007DecoherenceRevisited}:
\begin{quote}
It would be, after all, so much easier to believe in quantum physics if we could train our senses to perceive nonclassical superpositions. [...] There is, however, another reason for this
focus on the classical that must have played a decisive role: Our senses did not evolve for the purpose of verifying quantum mechanics. Rather, they have developed in the process in which survival of the fittest played a central role. 
\end{quote}
The present study of life expectancy for modes of experience exemplifies a way to turn such general ideas about fitness into explicit quantitative analysis.

\section{Experience does not encompass existence}\label{sec:ennee}

\subsection{Senses limit experiences}

In discussing physics and experience, it is important to note that physical world configurations are not in one-to-one correspondence with experiences. In fact much physical existence is not experienced, which means distinct physical configurations do not always correspond to distinct experiences.

\begin{figure}
    \centering
    \includegraphics[width=0.9\textwidth]{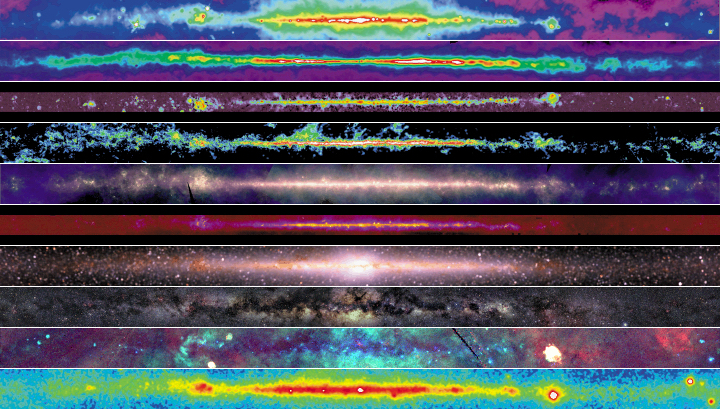}
    \caption{The Milky Way in different wavelengths.\protect\footnotemark}
    \label{fig:milky-way-spectrum}
\end{figure} 

\footnotetext{Image created by Jay Friedlander, reproduced with permission. Source: \textit{The Multiwavelength Milky Way} website of NASA's Goddard Space Flight Center: https://asd.gsfc.nasa.gov/archive/mwmw/mmw\_sci.html.}

For example, consider the Milky Way in the night sky. Throughout history, countless tales and poems have been devoted to its sheer beauty. Yet we only found out in the recent hundred years that what we appreciated is but a small fraction of the Galaxy's true physical existence (\Cref{fig:milky-way-spectrum}). A much wider spectrum exists beyond the optical one, but due to the limitations of our sense organ we could only miss them. Two physical configurations distinct in the non-optical spectrum does not give rise to distinct visual experiences for human beings. 


\begin{figure}
    \centering
    \includegraphics[width=0.5\textwidth]{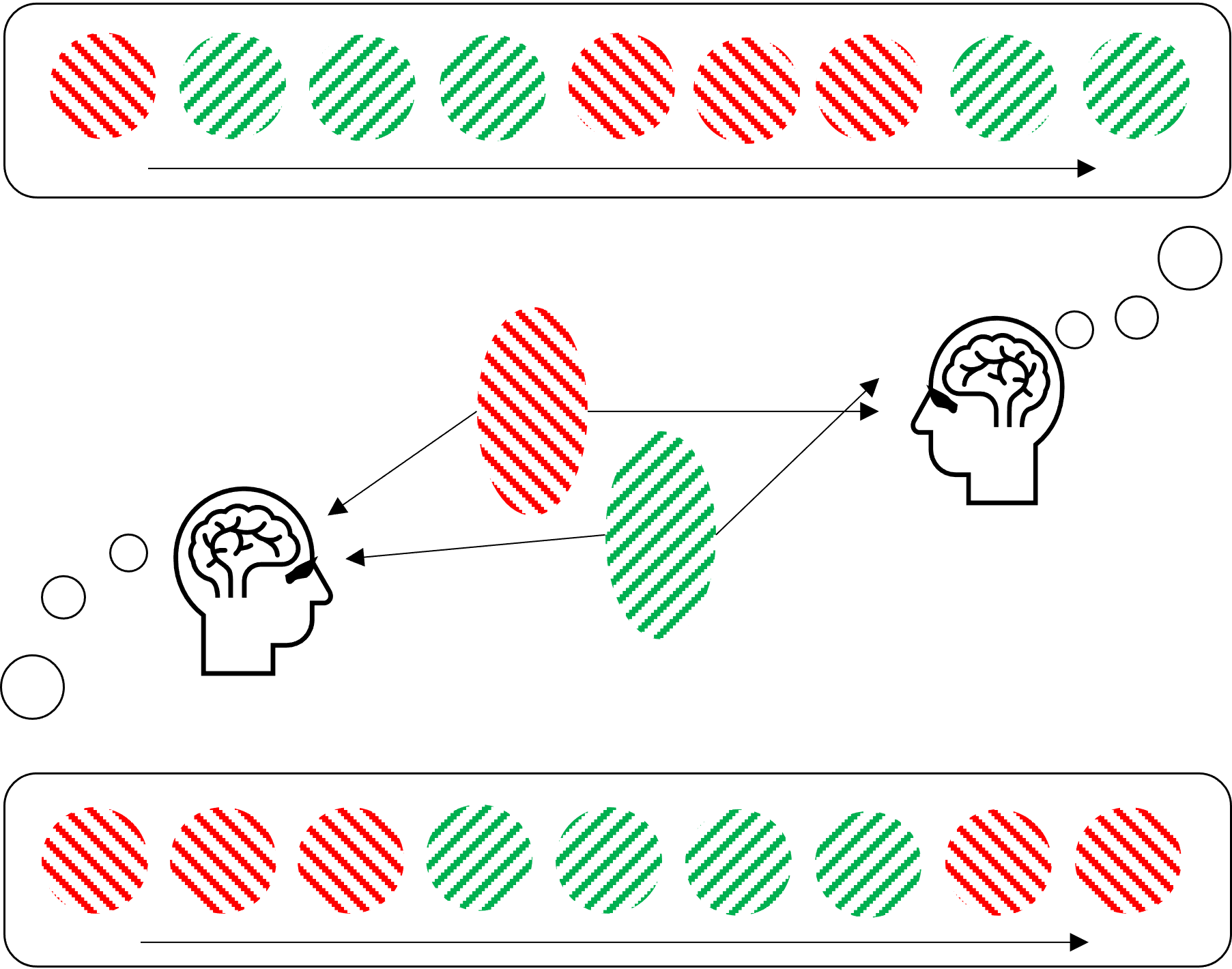}
    \caption{Binocular rivalry. When distinct images are shown to the two eyes, the images are perceived in alternation. Different individuals need not see the same view at the same time.}
    \label{fig:binocular}
\end{figure} 

\subsection{Multiplicities limit experiences}\label{sec:mle}

In the above example, physical existences evade experience because the sense organs are not built to perceive them. Yet on other occasions, physical existences evade experience even when they are totally perceptible by the sense organs. 

For example in the much studied phenomenon of binocular rivalry \cite{Blake2008BinocularRivalry}, a human being is presented with two very distinct images to the two eyes (\Cref{fig:binocular}). It turns out that at each time only one image is seen, while the other image drops out of visual perception even though it is right in front of the person's eye the whole time. Interestingly, the perceived view also switches stochastically between the two images with time durations conforming to gamma distributions. 

Such cases where multiplicity limits conscious experiences is actually the norm for human beings. At each moment, the sense organs register a great multitude of visual, auditory, somatosensory, olfactory, and gustatory physical sensory inputs. Yet only a small fraction of these produce conscious experiences which can be reflected upon and remembered \cite{Dehaene2014ConsciousnessThoughts}. Again, distinct physical configurations would not correspond to distinct experiences.

\subsection{Separating world and experiences}\label{sec:dwe}

\begin{figure}
    \centering
    \includegraphics[width=0.3\textwidth]{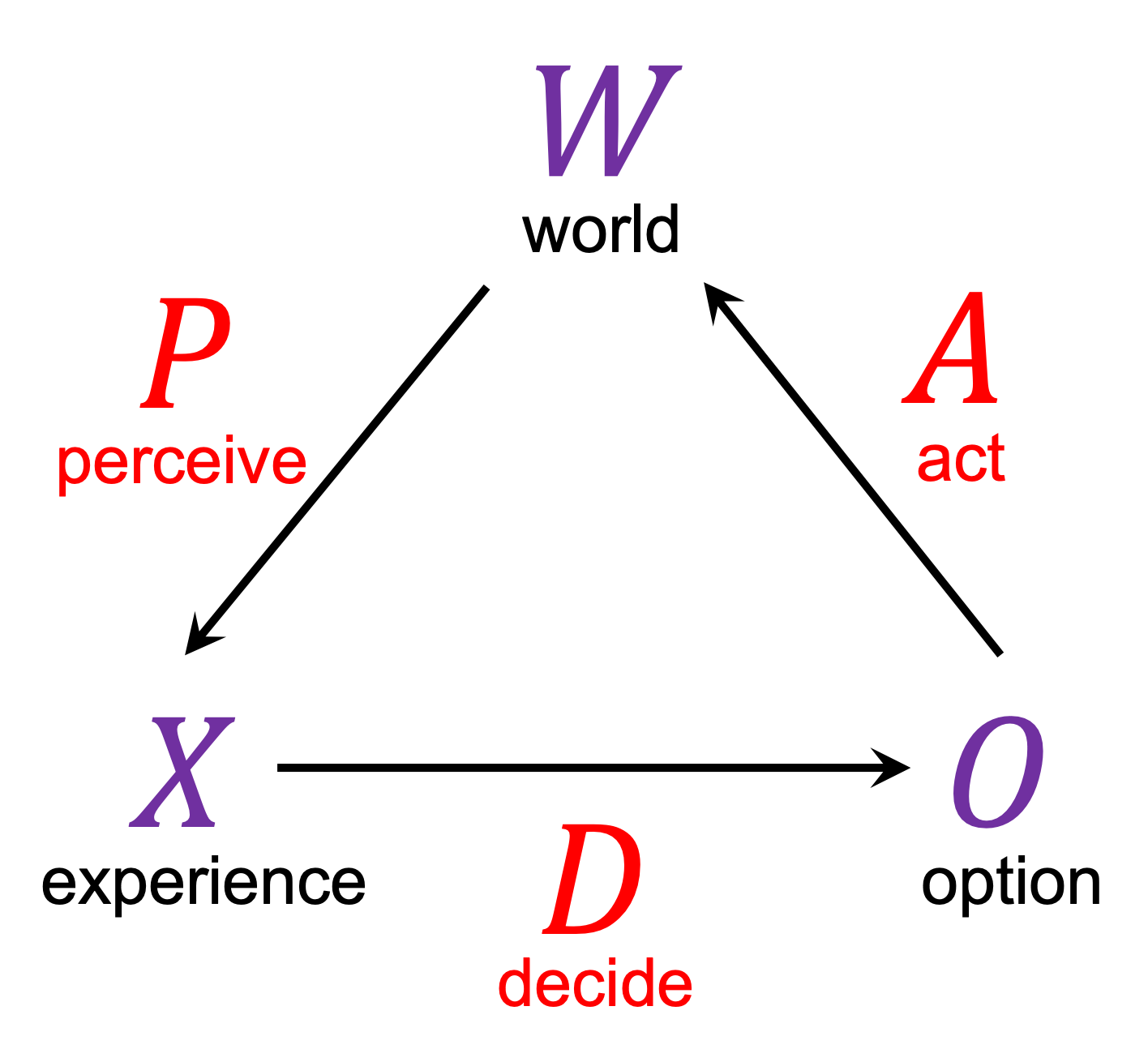}
    \caption{A adapted version of the perceive-decide-act loop of \cite{Hoffman2014ObjectsConsciousness}. The original ``action'' set $G$ is renamed into the ``option'' set $O$ to avoid name sharing with the  ``action'' set $A$.}
    \label{fig:pda_loop}
\end{figure} 

Since the physical world configurations are not in one-to-one correspondence with experiences, it helps to emphasis the distinction by treating them in separate sets. This is commonly done in psychological and cognitive studies. For instance, Hoffman \textit{et al.} \cite{Hoffman2014ObjectsConsciousness, Hoffman2015ThePerception} speak of the Perceive-Decide-Act (PDA) loop, which relate the distinct sets of world states $W$, experience states $X$, and options $O$ (\Cref{fig:pda_loop}).\footnote{I renamed the ``action'' set $G$ of the original work into the ``option'' set $O$ to avoid name sharing with the ``action'' set $A$.} In this regime, the perceptions of a being is characterized by elements in the experience set, instead of the physical configuration of the world set. An experience $x\in X$ prompts the being to decide on an option $o\in O$ for how to react with conditional probability $p(o|x)$. Based on that choice the being acts on the world to change it into the state $w\in W$ with conditional probability $p(w|o)$. That world state in turn is perceived to generate an experience $y\in X$ with conditional probability $p(y|w)$. 

The totality of the experiences of the being is characterized by a list of experience elements $x,y,z,\cdots \in X$ with conditional probabilities
\begin{align}\label{eq:1pe}
p(y|x)=\sum_{w,o} p(y|w)p(w|o)p(o|x)
\end{align}
relating sequential experiences. 

\section{Superposed world}\label{sec:sw}

\subsection{Objective world description}

To facilitate the study of ``modes of experience in the superposed world'' that follows, we specify precisely what ``superposed world'' means in this section. This is actually quite non-trivial if one wants a notion of the world that is objective. 

Consider again the example of binocular rivalry discussed above. Although two human beings viewing the same images will both experience alternating views conforming to a gamma-like distribution, the image they perceive at the same time need not be the same (\Cref{fig:binocular}). We may say that their visual perceptions are subjective and can differ from each other, while there is an objective world where the objects that give rise to the two images coexist all the time.

Suppose we want a notion of superposed world that is also objective. Then Wigner's friend setup \cite{Wigner1961RemarksQuestion} poses a problem. In this setting, Wigner assigns a state involving macroscopic superposition for the friend which is just another physical system, while the friend herself assigns a state where she is never in macroscopic superposition. If we were to use quantum states to describe the superposed world, it is not clear which state to use for an objective description.

\begin{figure}
    \centering
    \includegraphics[width=0.5\textwidth]{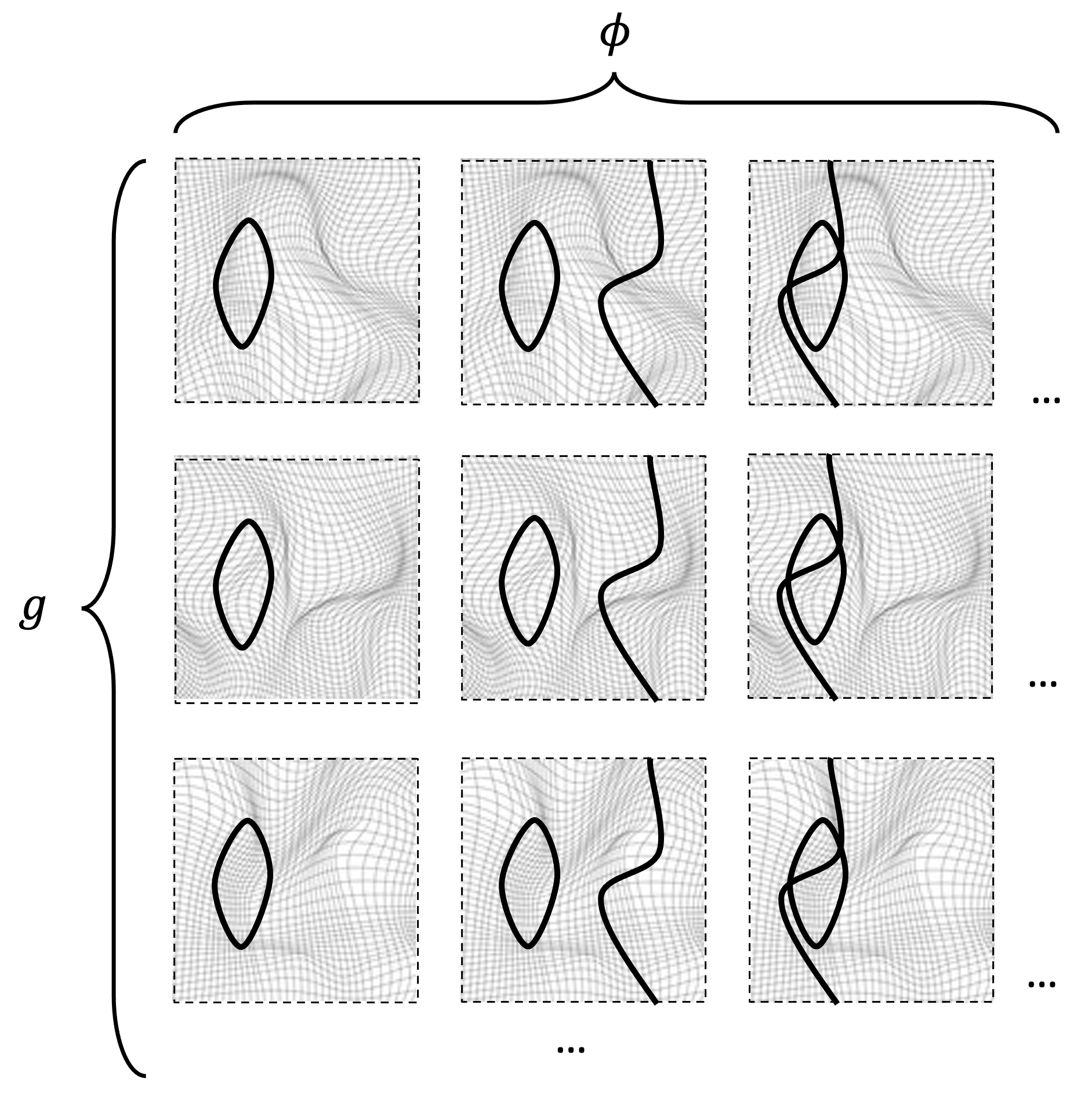}
    \caption{Gravity and matter configurations characterized by spacetime geometries $g$ and matter particle or field configurations $\phi$ living on them. Here $\Phi$ of \eqref{eq:pi1} consists of the joint matter gravity configurations $(g,\phi)$.}
    \label{fig:gmpi4}
\end{figure} 

The rest of this work makes use of the path integral formalism, which is the language to express the Standard Model for matter \cite{Donoghue2013DynamicsEdition} and most candidate theories for quantum gravity \cite{oriti2009approaches} (\Cref{fig:gmpi4}). In this language we have
\begin{align}\label{eq:pi1}
Z=\int D\Phi A[\Phi],
\end{align}
where $A[\Phi]$ is some function that incorporates possible boundary conditions and interior conditions. Here I refer to ordinary path integral boundary conditions that refer to the boundary part of the path integral configurations as ``\textbf{boundary conditions}'', and measurement weightings or other elements that refer to the interior part of the path integral configurations as ``\textbf{interior conditions}''. For instance, in computing the probability for an outcome of a sharp measurement taking place in the interior of the spacetime region of interest, we may want to restrict the path integral sum to only the subset of configurations compatible with the measurement outcome. In this case, the interior condition would be the characteristic function that assign $1$ to configurations compatible with some measurement outcome and $0$ to other configurations. If we have some boundary condition $\psi[\Phi]$ and interior condition $u[\Phi]$, then
\begin{align}\label{eq:ppi}
Z=\int D\Phi e^{iS[\Phi]} \psi[\Phi] u[\Phi]
\end{align}
where $A[\Phi]=e^{iS[\Phi]} \psi[\Phi] u[\Phi]$. In a Wigner's friend setup, Wigner and the friend would adopt different functions $A_W[\Phi]$ and $A_F[\Phi]$ for their different $\psi[\Phi] u[\Phi]$. 

There is an analogy with binocular rivalry. There the two objects that give rise to the two images are always simultaneously present, while human beings perceive only one image at a time for their subjective experiences. Here the path integral configurations $\Phi$ are always present, while subjective experience may differ. 
In the case of binocular rivalry, the objective world is taken to contain at the same time both objects. This suggests that we take as an objective account of the superposed world the set of all path integral configurations $\Phi$:
\begin{align}\label{eq:aos}
w=\{\Phi:\text{ path integral configurations}\}.
\end{align}
Even though different beings may assign different weight functions such as $A_W[\Phi]$ and $A_F[\Phi]$ in their own versions of \eqref{eq:pi1}, they all agree on configurations of $w$ as what is being superposed. In this sense $w$ forms the objective part of the superposed world.

\subsection{Flexibility}

The above description of the objective part of a superposed world is conservative, in the sense that more structures could hold in reality as objective for all experiential beings. For instance, the boundary condition for the universe $\psi[\Phi]$ of \eqref{eq:ppi} may be shared by the path integrals for all experiential beings. Or the universe may be fundamentally governed by an objective collapse model \cite{Bassi2013ModelsTests} so that all beings can in fact use a common objective physical description. 

If these turn out to be true, the description of the objective part of the superposed world \eqref{eq:aos} should be adjusted accordingly. Yet as long as the description is still in terms of a set that refers to multiple physical configurations in superposition, much of the consideration for modes of experience in the following section still applies, because it only needs the set structure of $w$ and does not care about the details of $w$.

\subsection{Beyond quantum}

The superposed world characterized by \eqref{eq:aos} goes beyond ``quantum superposition''. In addition to quantum superpositions with complex amplitudes, the superposed world characterized by \eqref{eq:aos} also supports other kinds of superpositions. For instance, in \Cref{sec:ex} I will present examples where all configurations are assigned the same amplitude, and examples where all configurations are assigned real amplitudes. The situation is reminiscent of General Probabilistic Theories (GPT) \cite{Popescu1994QuantumAxiom, HardyQuantumAxioms, barrett_information_2007} which go beyond quantum theories by allowing more general rules for probabilistic predictions. Here the focus is on more general modes of experiences, and the framework presented below may be said to support General Experience Theories (GET). This point will be elaborated on in \Cref{sec:fgpt}.

That \eqref{eq:aos} includes all configurations is also worth noting. Once the kinematical variables are set, \eqref{eq:aos} as the set of all path integral configurations essentially accommodates all conceivable physical configurations, whether one is considering quantum theory or some other theory. In fact, Lloyd and Dreyer have defined a ``universal path integral'' that encompasses all computable structures \cite{Lloyd2015TheIntegral}, which highlights how general the path integral configurations set could be. 

\section{Modes of experience}\label{sec:moe}


\subsection{The World is One. The One is All.}

The superposed world in \Cref{sec:sw} is fairly special. In order to isolate the objective part of the world, we stripped away the amplitudes. The set of world states
\begin{align}
W=\{w\}
\end{align}
contains just \textit{one} state $w$ of \eqref{eq:aos}. The element $w$ in turn accommodates \textit{all} path integral configurations. In this sense, the superposed world is One, and the One is All.




The superposed world is special in that it never changes! This poses an obstacle to apply the PDA-loop of \Cref{sec:dwe} to relate the world with experiences. Because the world has just one state $w$, the act map of \Cref{sec:dwe} obeys $p(w|o)=1$ for all options $o$, indicating that the world stays the same no matter what action is performed. Consequently, the first person experience probabilities \eqref{eq:1pe} for the PDA-loop of \Cref{sec:dwe} reduce according to
\begin{align}
p(y|x)=&\sum_{w,o} p(y|w)p(w|o)p(o|x)
\\
=&\sum_{w,o} p(y|w)p(o|x)
\\
=&\sum_{w} p(y|w)
\\
=&p(y),
\end{align}
where in the last line I noted that there is just one $w$ so omitted it in $p(y|w)$. That $p(y|x)=p(y)$ means the probabilities for the next experiences $y$ exhibit no correlation with the previous experiences $x$. Therefore new schemes are needed to relate the world, the experiences, and the options if it is to incorporate sequential experiences that exhibit correlations.




\subsection{Modes of experience}

In principle, beings conforming to different schemes relating the world, the experiences, and the options could coexist in the same world. Therefore instead of fixing on one scheme, we consider multiple schemes.

\begin{figure}
    \centering
    \includegraphics[width=0.9\textwidth]{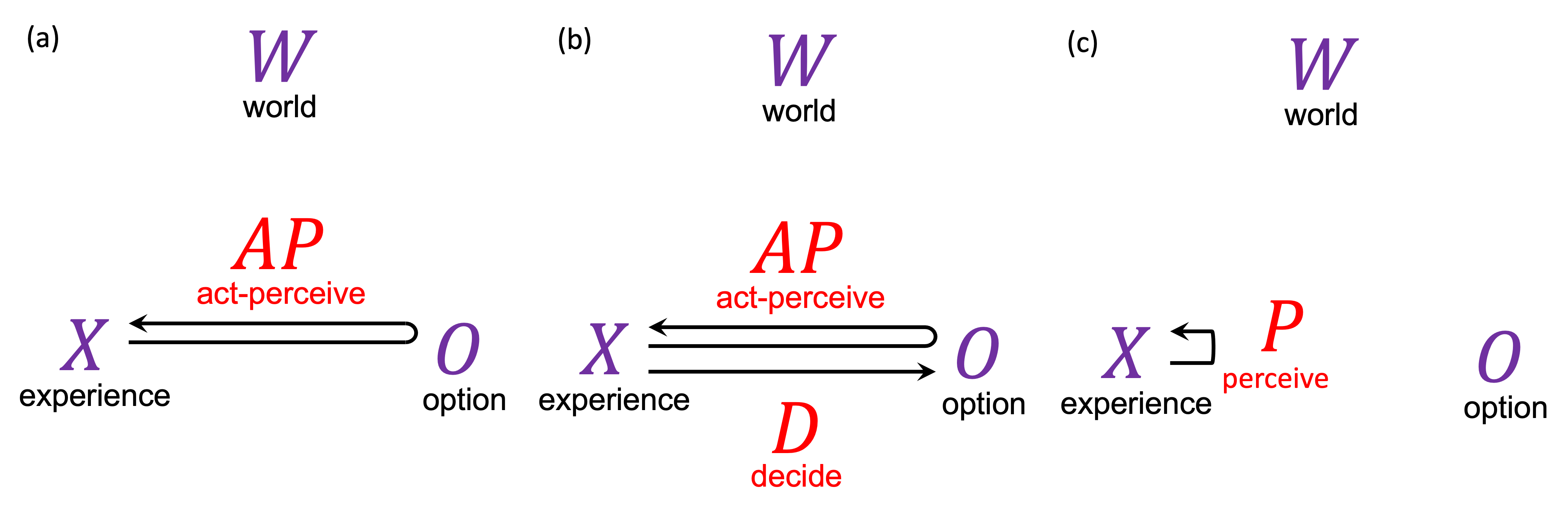}
    \caption{(a) AP scheme; (b) D-AP scheme; (c) P scheme}
    \label{fig:pda_schemes}
\end{figure} 

Three examples are shown in \Cref{fig:pda_schemes}, with the following probabilistic rules relating sequential experiences.
\begin{example}[AP scheme]
Act-perceive map: $p(y|x,o)$.
\end{example}
Here an experience $x$ together with an option $o$ give rise to another experience $y$ with probability $p(y|x,o)$. This scheme may possibly be developed to incorporate Libertarian Free Will, because the options are not determined (probabilistically or deterministically) by any other variable. 

\begin{example}[D-AP scheme]
Decide map: $p(o|x)$. Act-perceive map: $p(y|x,o)$. Experience map: $p(y|x)=\sum_o p(y|x,o)p(o|x)$.
\end{example}
Here an experience $x$ leads to an option $o$ with probability $p(o|x)$. The experience $x$ and option $o$ then jointly leads to an experience $y$ with probability $p(y|x,o)$. The probability rule relating sequential experiences is encoded in $p(y|x)$.

\begin{example}[P scheme]
Perceive map: $p(y|x)$.
\end{example}
Here options play no role. An experience $x$ probabilistically determines the next experience $y$ directly.

These examples are not meant to be exhaustive. In general, a \textbf{mode of experience} in a superposed world with the configuration set $W$ consists of:
\begin{itemize}
\item An experience set $X$, plus an option set $O$ if applicable.
\item A perceive, decide, act scheme diagram relating the world, the experiences and the options.
\item Probabilistic rules for correlated experiences, e.g., in the form of $p(y|x,o)$, $p(y|x)$ etc. for $x,y\in X$ and $o\in O$
\end{itemize}
A mode of experience with these ingredients may be viewed as a predictive theory of experience. The quantities $p(y|x,o)$, $p(y|x)$ etc. yield probabilistic predictions for subsequent experiences given previous experiences.

\subsection{Candidate modes of experience}\label{sec:cme}

Specifying the experience set $X$ turns out to be difficult in practice. One needs to first find a language to express the experiences, and then give an exhaustive list of all possible experiences in this mode. 

One possible language is that of computer science, where one uses bit strings to encoded experiences. This route is taken, for instance, by Markus M{\"u}ller in his algorithmic information theory approach to experience and physics \cite{Muller2020LawTheory}. Another possible language is that of physics, where one uses physical configurations to describe experiences. In this work, we adopt physicists' language and focus on physical configurations. Ultimately the physical description can be converted to bit strings, so this choice is not a fundamental one. It just says we focus on the physical configuration level of description.


The difficulty in describing experiences in terms of physical configurations is that first, there is no guarantee that all experiences are describable in terms of physical configurations. Furthermore, even for experiences describable in terms of physical configurations, it is not clear what the correspondence map is. 


To make progress despite these difficulties, I make a concession. Instead of specifying the set of experiences $X$ in the language of physical configurations, we deal directly with some set of expressions in terms of physical configuration. In this \textbf{condition set} $C$, the elements are expressions of physical configuration that form our best beliefs about the physical conditions for experiences. For instance some of them could be in terms of the various physical brain configurations of human beings which is believed to correspond to various experiences. 


The concession is that we do not know for sure if all elements in $C$ correspond to genuine experiences in $X$, or if all experiences in $X$ have corresponding elements in $C$. By working with physical conditions in $C$ instead of genuine experiences in $X$, we defer the important but difficult task to tell the exact relationship between $X$ and $C$ to the future.

Replacing the experience set $X$ of a mode of experience by the condition set $C$, we obtain a framework for \textbf{candidate modes of experience}, which only form candidates for modes of experience in view of the uncertainty about how $C$ relates to $X$. In a superposed world with the configuration set $W$, a candidate mode of experience consists of:
\begin{itemize}
\item A condition set $C$, plus an option set $O$ if applicable.
\item A perceive, decide, act scheme diagram relating the world, the conditions and the options.
\item Probabilistic rules for correlated experiences, e.g., in the form of $p(d|c,g)$, $p(d|c)$ etc. for $c,d\in C$ and $o\in O$
\end{itemize}

Given the uncertainty on how $C$ relates to $X$, an element $c\in C$ may be said to describe a candidate experience. In comparison to modes of experience, adopting physical configurations to describe candidate experiences puts the world back into the picture. In a mode of experience with the scheme diagrams of \Cref{fig:pda_schemes}, the world set $W$ completely decouples from the experience and option sets. In contrast, in candidate modes of experience the candidate set $C$ refers to physical configurations of the world set element $w$. This connects $W$ to $C$.

\subsection{Examples}\label{sec:ex}

\begin{figure}
    \centering
    \includegraphics[width=0.9\textwidth]{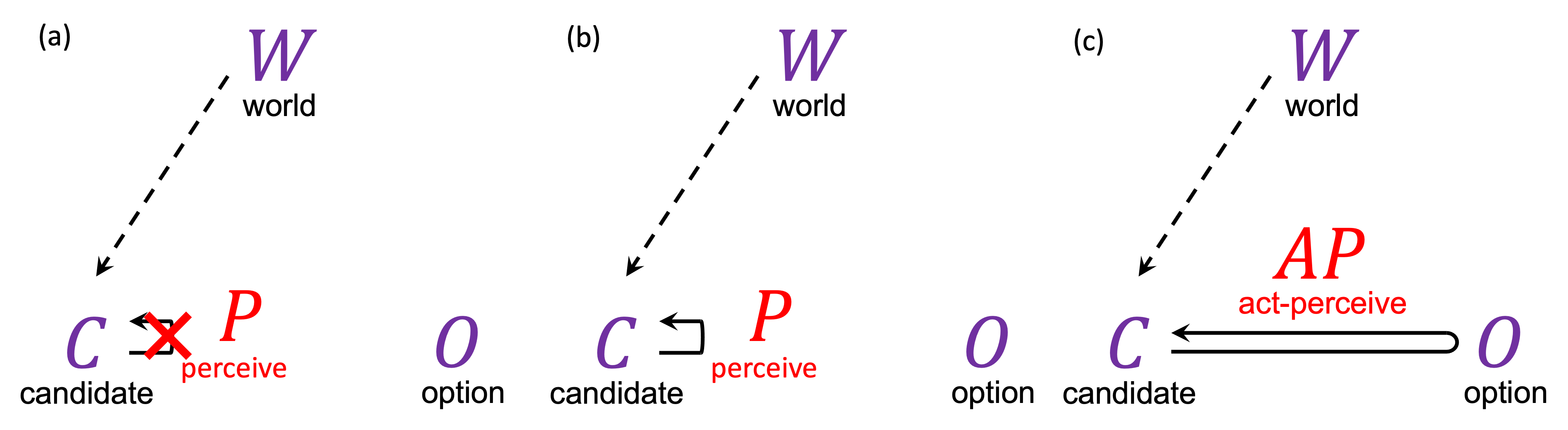}
    \caption{(a) Scheme diagram for <Omniscient> and <Ephemeral> whose perceptions lack sequential correlations; (b) Scheme diagram for <Deterministic>, <Quantum-1> and <Real-Quantum-1>; (c) Scheme diagram for <Quantum-2>.}
    \label{fig:pda_schemes_2}
\end{figure} 

Let us consider some examples of candidate modes of experience with scheme diagrams of \Cref{fig:pda_schemes_2}. Here I adopt the convention ``<Name>'' in assigning names to candidate modes of experience.
\begin{example}[<Omniscient>]
$C\simeq W=\{w\}$. $p(w)=1$.
\end{example}
In <Omniscient>, the condition set $C$ is isomorphic to the world set $W$. The only element $w$ is the set of all physical configurations. A being in this mode experiences all there is in the world, so is ``omniscient''. 

The experiences we are familiar with are sequential ones. Yet in <Omniscient> only one experience corresponding to $w\in C$ is realized. This demonstrates the possibility to consider modes of non-sequential experiences in the present language.

\begin{example}[<Ephemeral>]
Arbitrary $C$. $p(d|c)=p(d)$.
\end{example}
In <Ephemeral>, sequential experiences $c$ and $d$ exhibit no correlations.\footnote{See \cite{PageSensibleProbabilistic, Page1995SensibleMind, Page2003MindlessConsciousness} for a quantum formalism for sentient experiences that also does not refer to correlations between experiences.} 
The experience of a being with no memory system may conform to this mode.

\begin{example}[<Deterministic>]
Arbitrary $C$. $p(d|c)=\delta(d,d_c)$.
\end{example}
In <Deterministic>, the next experience $d$ is uniquely determined from the previous experience $c$ to be $d_c$, and the experiences are always realized with probability one. Together with <Omniscient>, it shows that although the framework refers to probabilities for experiences, it incorporates deterministic experiences as special cases.

\begin{example}[<Quantum-1>]
The condition set $C\subset\{c\mid c:w\rightarrow \mathbb{C}\}$ is some subset of the complex functions on $w$, and
\begin{align}\label{eq:quantum1}
p(d|c)=&\frac{\abs{Z[c,d]}^2}{\sum_d \abs{Z[c,d]}^2},
\\
Z[c,d]=&\int D\Phi ~ A[\Phi] ~ \psi[\Phi] ~ c[\Phi] ~d[\Phi].\label{eq:quantum2}
\end{align}
Here $c,d\in C$ are conditions for two sequential experiences, $\psi[\Phi]\in \mathbb{C}$ is the boundary condition, and $A[\Phi]\in \mathbb{C}$ is the amplitude map which usually takes the form $A[\Phi]=e^{iS[\Phi]}$ in a path integral.
\end{example}

The precise form of the condition set $C$ is currently unknown, and it depends on what functions $c$ correspond to experiences. For ordinary beings it seems reasonable to assume that the experiences are localized to quantum or classical spacetime regions. If an experience is localized to the spacetime region $R$, then one could assume that the condition $c$ for this experience obeys $c[\Phi_1]=c[\Phi_2]$ whenever two configurations $\Phi_1,\Phi_2\in w$ agree in region $R$. 

The name <Quantum-1> has a number in it because this is not the only possible mode of experience that one would consider as quantum. For example, another mode is given in the next example that differs from this one by getting actions involved. In addition, one could generalize the path integral to a double path integral to accommodate mixed boundary conditions and weights.


\begin{figure}
    \centering
    \includegraphics[width=0.5\textwidth]{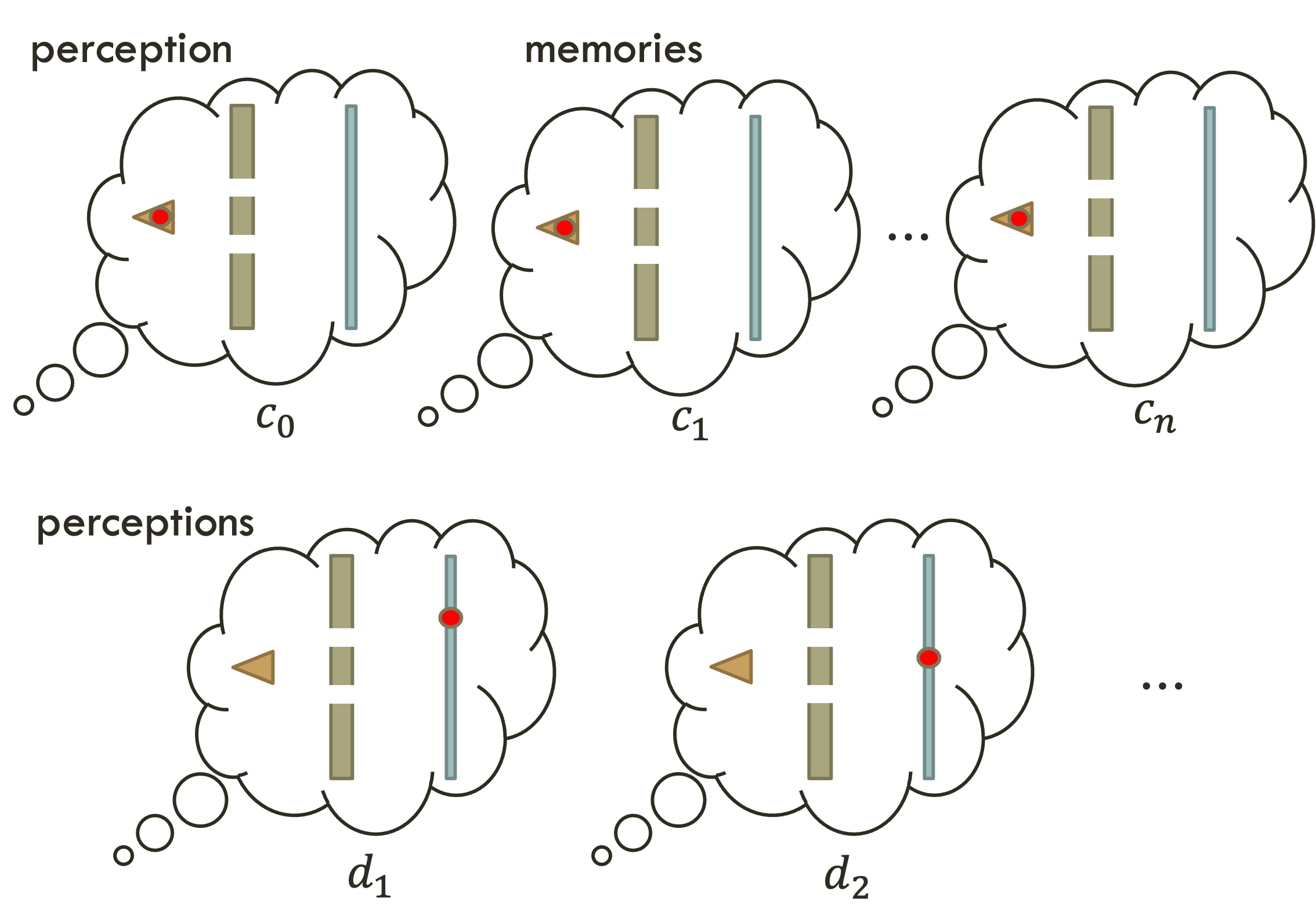}
    \caption{In <Quantum-1>, the description of a double-slit experiment refers to a series of experiences of a being consisting of both sensory perceptions and memories.}
    \label{fig:double-slit}
\end{figure} 

To illustrate how <Quantum-1> describes ordinary quantum experiments let us consider a double-slit experiment. As illustrated in \Cref{fig:double-slit}, the first person experience of the experimenter consists of a series of sensory perceptions and memories. In the beginning, the person perceives the experimental setup where a particle is ready to be fired towards the screen on the other side of the slit. This perception corresponds to a condition $c_0\in C$. The total initial experience of the person corresponds to $c_O = c_0\bar{c_0}\in C$, which is joined by $\bar{c_0}$ that provides the rest condition for the person's experience (such as having a brain configuration of excitement in performing the double-slit experiment). The element $\bar{c_0}$ encompasses all the other ingredients of the physical condition that the next experience conditions on, so can actually contain a lot of content.

Next, the person looks away from the particle for some duration of time. There is a sequence of conditions $c_I = c_1\bar{c_1},c_{II} = c_2\bar{c_2},\cdots, c_N=c_n\bar{c_n} \in C$, where the $c_i$'s are conditions for having the memories of seeing the experimental setup and the $\bar{c_i}$'s are the conditions for the rest of the person's experiences. The conditional probabilities for adjacent candidate experiences are given by  \eqref{eq:quantum1} and \eqref{eq:quantum2}. For instance, after $c_O$, the probability for $c_I$ is
\begin{align}
p(c_I|c_O)=&\frac{\abs{Z[c_I|c_O]}^2}{\sum_{c_I\in C_{I}} \abs{Z[c_I|c_O]}^2},
\\
Z[c_I|c_O]=&\int D\Phi ~ A[\Phi] ~ \psi[\Phi] ~ c_O[\Phi] ~c_I[\Phi],
\end{align}
where $C_{I}$ is the set of all possible conditions that follow $c_O$. Concretely, $A[\Phi]$ could be $e^{iS[\Phi]}$ with $S$ as the action as in ordinary path integrals, $\psi$ could be a boundary condition for the universe, and $c_O, c_I$ could be characteristic functions that assign $1$ to $\Phi$ compatible with the designated brain configurations for the person and $0$ to otherwise.

After this duration of time, the person looks at the screen. There will be a list of possible perceptions for seeing the record of the particle landing at different locations. These correspond to the conditions $d_1,d_2,\cdots$. Again the conditional probabilities are computed using \eqref{eq:quantum1} and \eqref{eq:quantum2}. The suitably chosen functions $c_N,\bar{c_N},d_i,\psi$ would allow two interfering dominant contributions to the path integral corresponding to the particle paths through the two slits. This should yield the expected probability distribution $p(d_i|c_N)$ for the double-slit experiment. 

The above description of the double-slit experiment is close to that of an ordinary path integral. The essential difference is that we focus on first person experiences of the process and refer explicitly to the brain configurations of the person through the $c$'s and $d$'s in the path integrals.

The candidate mode <Quantum-1> still follows the scheme of \Cref{fig:pda_schemes_2} (b) where options are not involved. As a variant of <Quantum-1> we have:
\begin{example}[<Quantum-2>]
$C\subset\{c\mid c:w\rightarrow \mathbb{C}\}$, and
\begin{align}
p(d|c,o)=&\frac{\abs{Z[c,d,o]}^2}{\sum_d \abs{Z[c,d,o]}^2},
\\
Z[c,d,o]=&\int D\Phi ~ A[\Phi] ~ \psi[\Phi] ~ c[\Phi] ~d[\Phi]~\chi[d,o],
\end{align}
where $o$ belong to the option set $O$.
\end{example}
<Quantum-2> follows the scheme of \Cref{fig:pda_schemes_2} (c), where the option set $O$ is relevant. An element $o\in O$ affects the probabilities $p(d|c,o)$ through the functions $\chi[d,o]$ that go into the path integral formula. For instance, if the choice of $o$ disallows a certain perception $d$, this is modelled by $\chi[d,o]=0$.

In addition to complex Hilbert spaces, quantum theory has also been  considered on real Hilbert spaces \cite{Stueckelberg1960QuantumHilbert-Space}, where the theory exhibits interesting properties such as the violation of local tomography \cite{Hardy2010LimitedTheory}. In the present framework there goes the candidate mode <Real-Quantum-1> which assigns real amplitudes to the path integral configurations:
\begin{example}[<Real-Quantum-1>]
$C\subset\{c\mid c:w\rightarrow \mathbb{R}\}$, and
\begin{align}
p(d|c)=&\frac{\abs{Z[c,d]}^2}{\sum_d \abs{Z[c,d]}^2},
\\
Z[c,d]=&\int D\Phi ~ A[\Phi] ~ \psi[\Phi] ~ c[\Phi] ~d[\Phi],
\label{eq:rqpf}
\end{align}
where $c,d\in C$, and $\psi[\Phi],A[\Phi]\in \mathbb{R}$ are \textit{real-valued} functions.
\end{example}
One example for a real $A$ is $A[\Phi]=e^{-S[\Phi]}$ where $S$ is a real action. Just like <Quantum-1> has variants such as <Quantum-2>, <Real-Quantum-1> has variants that for instance introduces decisions into the scheme.

\section{Evolutionary considerations}\label{sec:ec}

\subsection{Outline}

Not so long ago, it was common to believe that human beings have always been the way they are. Evolutionary biology changed that. We now believe that in the long run the biological forms can undergo drastic changes shaped by natural selection. This offers a way to explain why organisms behave the way they do and exhibit features that they possess.
If alternative modes of experience exist in the universe, then it is natural to ask if modes of experience can evolve. This could yield explanations of why we experience the way we do.



In this section, I study life expectancy as an example of an evolutionary fitness function for some different candidate modes of experience in a toy universe of 1D scalar field theory. It is shown that <Quantum-1> without macroscopic superposition has longer life expectancy than <Quantum-1> with macroscopic superposition, while both have longer life expectancies than <Real-Quantum-1>.

These results admit some simple explanations. Since <Real-Quantum-1> is based on the Euclidean action, the amplitudes of ``excited'' (alive) configurations quickly decay to death configuration due to the exponential suppression by the real exponent for the path integral amplitude. Since <Quantum-1> with macroscopic superpositions can put an alive configuration into superposition with another alive configuration that is dynamically closer to the death configuration, it shortens the life expectancy in comparison to <Quantum-1> without macroscopic superpositions.

\subsection{Life expectancy}\label{sec:le}

For a mortal being, there are death configurations represented by conditions $D$. For beings that do not revive,
\begin{align}
p(c|D)=0
\end{align}
for all conditions $c$ for living experiences. 

Given a set of death configurations in a candidate mode of experience, it is not hard to use the conditional probabilities $p(d|c)$ to compute the life expectancy of beings in this mode. For simplicity, let us consider a scheme like \Cref{fig:pda_schemes_2} (b) where options play no role. Suppose the condition set $C$ is finite. Then the perception map $p(d|c)$ forms a stochastic matrix, with $c\in C$ as the column index and $d\in C$ as the row index. In order to compute the life expectancy, we take away the columns and rows of the death configurations from $p(d|c)$ to obtain a \textbf{reduced matrix} $T$, which encodes the transition probabilities between alive configurations. 

Suppose initially the being is in the alive configurations $c$ with probabilities $p(c)$, and collect these in the vector $v$ with components $v_c=p(c)$. Then at the first time step, the being is alive with probability $\norm{v}_1$, where $\norm{v}_1=\sum_c v_c$ is the 1-norm. At the second time step, the being is alive with probability $\norm{Tv}_1$ etc. In general, at the $s$-th time step, the being is alive with probability
\begin{align}\label{eq:pas}
q(s):=\norm{T^{s-1}v}_1.
\end{align}

Let $q(s+1|s)$ be the conditional probability that the being is alive at step $s+1$ when it is alive at step $s$, and let $p_{\text{age}}(s)$ be the probability for the being to live till step $s$ and die at step $s+1$. Then
\begin{align}\label{eq:pac}
q(s+1)=&q(s+1|s)q(s),
\\p_{\text{age}}(s)=&q(s)(1-q(s+1|s))
\\=&q(s)-q(s+1).\label{eq:page}
\end{align}
The life expectancy of the being is given by 
\begin{align}
\ev{s}=&\sum_s p_{\text{age}}(s) s
\\=&(q(1)-q(2))+2(q(2)-q(3))+3(q(3)-q(4))+\cdots,
\\=&q(1)+q(2)+\cdots
\\=&\norm{(I+T+T^2+\cdots)v}_1
\\=&\norm{(I-T)^{-1}v}_1,\label{eq:le}
\end{align}
where \eqref{eq:pas} is used in the second to last line. Therefore given the initial probability vector $v$, the life expectancy can be computed from $T$, the submatrix of $p(d|c)$ for the living configurations.

\subsection{Scalar field theory}

We want to compare the life expectancies for some candidate modes of experience. For concreteness and simplicity, we work with a scalar field theory instead of the more realistic Standard Model-Quantum Gravity coupled path integral, and discuss in \Cref{sec:bsm} the prospects for more advanced models.

\begin{figure}
    \centering
    \includegraphics[width=0.5\textwidth]{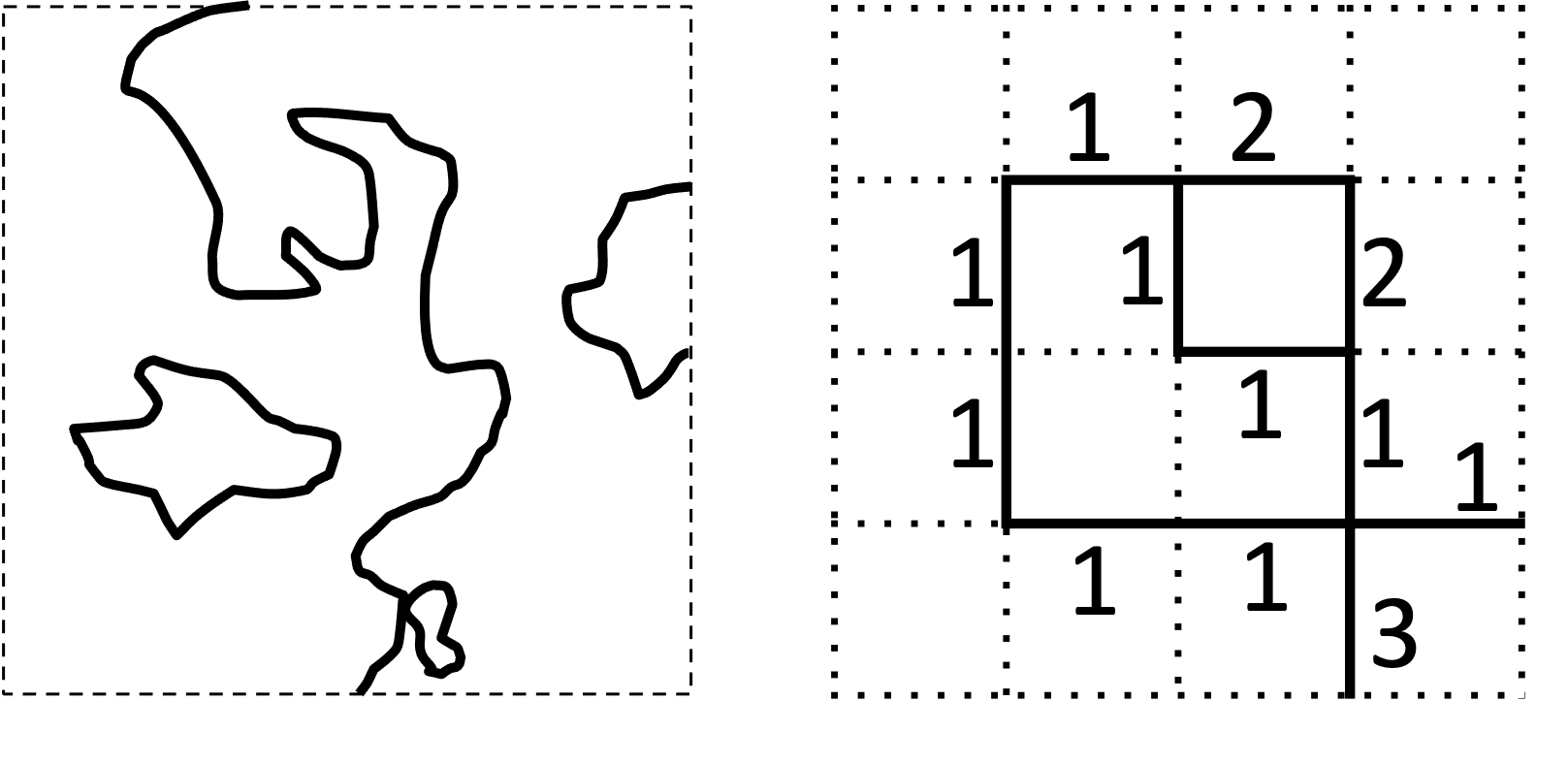}
    \caption{Left: particle configurations for scalar field theory. Right: lattice particle configurations.}
    \label{fig:pwl}
\end{figure} 

As shown in \Cref{sec:sfpi}, the path integral of a real scalar field also admits a particle representation after performing a simple Taylor series expansion. In this representation, a configuration consists of a set of particle worldlines, and the path integral sums over all such configurations where the number of worldlines is arbitrary (\Cref{fig:pwl}). On a lattice, a configuration is encoded in a list of integers
\begin{align}
\vec{n}=\{n_e\}_e,
\end{align}
where $n_e$ represents the number of worldline segments crossing the lattice edge $e$.

From \Cref{sec:sfpi}, the partition function takes the form
\begin{align}\label{eq:rsZ}
Z=& \mathcal{N}\sum_{\vec{n}\text{ extended}} \prod_{e} E_e(n_e) \prod_{v} V_v(n_v),
\end{align}
where $\mathcal{N}$ is a constant. As explained in \Cref{sec:sfpi}, in a globally $Z_2$-symmetric theory that we consider here, only extended particle configurations are included in $\sum_{\vec{n}\text{ extended}}$. This means that a particle line entering an interior vertex $v$ must extend beyond it. Denote by 
\begin{align}
n_v=\sum_{e\in v} n_e
\end{align} 
the total number of particles crossing the vertex $v$. Then in an extended configuration, 
\begin{align}
n_v\text{ is even at all interior vertices }v,
\end{align}
while $n_v$ can also be odd on the boundary of the region of spacetime under consideration since an extended particle line may end there. 


The edge amplitude at $e$ and vertex amplitude at $v$ are
\begin{align}
E_e(n_e)=&\begin{cases}
\frac{(-i)^{n_{e}}}{n_{e}!}, \quad \text{<Quantum-1>}
\\
\frac{1}{n_{e}!}, \quad \text{<Real-Quantum-1>}
\end{cases}
\\
V_v(n_v)=&\begin{cases}\label{eq:vvn}
2\int_0^\infty d r ~ r^{n_v}e^{-i\eta r^2-iV(r)}, \quad \text{<Quantum-1>}
\\
2\int_0^\infty d r ~ r^{n_v}e^{-\eta r^2-V(r)}, \quad \text{<Real-Quantum-1>}
\end{cases}
\end{align}
where $\eta$ is the renormalized mass parameter, related to the spacetime dimension $D$, the bare mass $m$ and the lattice spacing $a>0$ by
\begin{align}\label{eq:eta}
\eta &=
\begin{cases}
a^2 m^2/2+D-2, \quad \text{<Quantum-1>}
\\
a^2 m^2/2  + D, \quad \text{<Real-Quantum-1>}.
\end{cases}
\end{align}

\subsection{Probability formula}\label{sec:pf}

In the present context, candidate experiences are described by particle line configurations. For instance, suppose the brain of a being exists in a region $R$ of spacetime. Then a subregion $R_1\subset R$ can have a particle configuration $m=\{n_e\}_{e\in R_1}$ describing one candidate experience, while an adjacent subregion $R_2\subset R$ can have another particle configuration $n=\{n_e\}_{e\in R_2}$ describing another candidate experience. 

In <Quantum-1> and <Real-Quantum-1>, the probability amplitude for sequential experiences are given by
\begin{align}\label{eq:Anm1}
A(n|m)=\sum_{\vec{n}} A[\vec{n}] \psi[\vec{n}] \delta[\vec{n}_{R_1},m] \delta[\vec{n}_{R_2},n].
\end{align}
Here $\psi$ is some boundary condition, and the conditions are delta functions which enforce that $\vec{n}$ agree with $m,n$ in regions $R_1,R_2$. $A[\vec{n}]$ is fixed by the summand of \eqref{eq:rsZ} so that
\begin{align}\label{eq:Anm2}
A(n|m)=\mathcal{N}\sum_{\vec{n}\text{ extended}} \prod_{e} E_e(n_e) \prod_{v} V_v(n_v) \psi[\vec{n}] \delta[\vec{n}_{R_1},m] \delta[\vec{n}_{R_2},n].
\end{align}
The conditional probability for sequential candidate experiences is then
\begin{align}\label{eq:cp}
p(n|m)=\frac{\abs{A(n|m)}^2}{\sum_n \abs{A(n|m)}^2}.
\end{align}
Since $\mathcal{N}$ of \eqref{eq:Anm2} gets cancelled out in \eqref{eq:cp}, I will omit it in the following.

\begin{figure}
    \centering
    \includegraphics[width=0.5\textwidth]{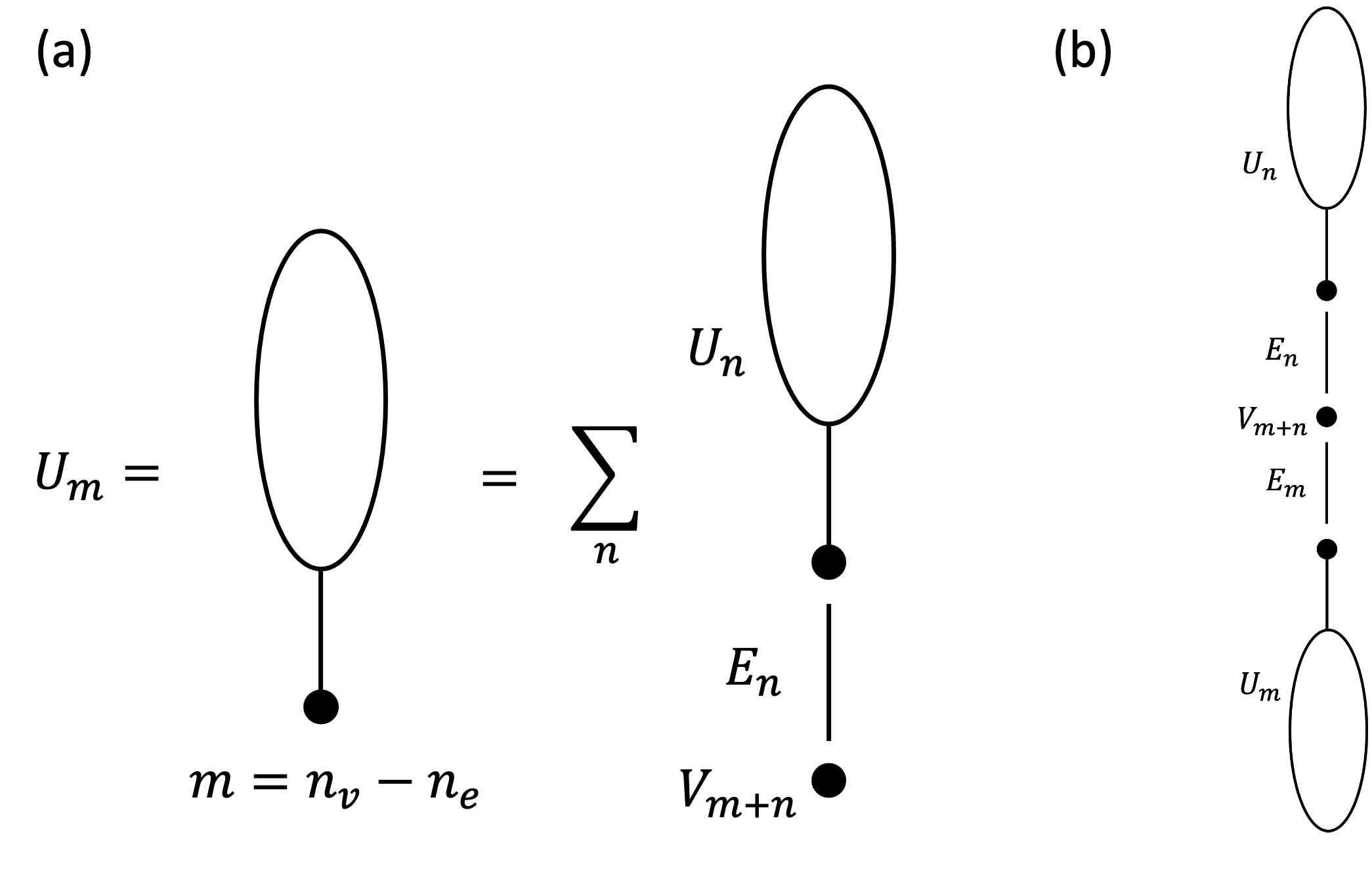}
    \caption{(a) Concatenate a configuration in $U_n$ with an $n$-edge and a $m+n$-vertex. This yields back a configuration in $U_m$. All configurations in $U_m$ can be obtained this way for some $n$, so summing over $n$ yields $U_m$. (b) Sandwiching $E_{n} V_{m+n} E_m$ between $U_n$ and $U_m$ yields the sum over all configurations with $m$ particles on one edge and $n$ particles on the adjacent edge.}
    \label{fig:Ti2}
\end{figure} 

The task now is to compute \eqref{eq:cp} explicitly. 
For simplicity, I specialize to an 1D spacetime without boundary and consider sequential candidate experiences localized to adjacent individual edges. In this simple model, an integer $n_e$ on the edge $e$ can be thought of as a brain configuration giving rise to some particular experience localized to this edge. 
Furthermore, since the spacetime is unbounded, the boundary condition is trivializes into
\begin{align}
\psi=1.
\end{align}
Consequently an extended configuration must have even $n_e$ on all edges.

To obtain a nicer formula, we consider a half-bounded region of 1D spacetime, with $v$ as the boundary vertex and $e$ the edge  adjacent to it. For any fixed integer $m\ge 0$, consider the set of all particle configurations so that $n_v-n_e=m$ (\Cref{fig:Ti2} (a)). Denote by $U_m$ the amplitude sum over this set, by $E_n$ the edge amplitude with $n_e=n$ for arbitrary $n$, and by $V_{n}$ the vertex amplitude with $n_v=n$ for arbitrary $n$. Then  (\Cref{fig:Ti2} (a))
\begin{align}\label{eq:Ti}
    U_m=\sum_{n} U_n  E_n V_{m+n}.
\end{align}
In terms of $U_m$ we have (\Cref{fig:Ti2} (b))
\begin{align}\label{eq:anm}
A(n|m)=U_n E_{n} V_{m+n} E_m U_m.
\end{align}
Plugging this in \eqref{eq:cp} yields
\begin{align}\label{eq:cp1}
p(n|m)=\frac{\abs{U_n E_{n} V_{m+n}}^2}{\sum_n \abs{U_n E_{n} V_{m+n}}^2},
\end{align}
where the common factor $\abs{E_m U_m}^2$ dropped out.

As mentioned, in the unbounded spacetime all edges must have even $n_e$. Therefore $m$ and $n$ only assume non-negative even integer values in \eqref{eq:Ti}. Define the operator $M$ and vector $u$ by the particle  number basis elements
\begin{align}\label{eq:Mij}
M_{i,j}=&V_{2i+2j-4} E_{2j-2}
\\u_i=&U_{2i-2}.\label{eq:uU}
\end{align}
Then \eqref{eq:Ti} translates to
\begin{align}\label{eq:uMu}
u_i=\sum_j M_{i,j} u_j.
\end{align}
Since $M$ is given in terms of $E$ and $V$, we can solve \eqref{eq:uMu} for its eigenvector with eigenvalue $1$ to obtain $u$. This in turn encodes all the elements $U_m$ through \eqref{eq:uU}, which can be used in \eqref{eq:cp1} to obtain $p(n|m)$.

\subsection{Settings}

A few further settings are needed in order to carry out numerical computations.
First, in computing $p(n|m)$ we need to decide the range of $m$ and $n$. The theory itself allows all non-negative integer values for $m$ and $n$, but in practice we can only compute finitely many values in a numerical computation. Therefore we cutoff at $N$ distinct particle numbers so that $m,n=0,1,2,\cdots,N-1$ and $p(n|m)$ is a stochastic matrix with $N\times N$ entries. For concreteness we perform calculations for
\begin{align}
N=5,9,13.
\end{align}

Next, we need to fix the potential $V$ in the action. Here I consider the free theory with $V=0$ for \eqref{eq:vvn} so that
\begin{align}
V_v(n_v)=&\begin{cases}
(i\eta)^{-\frac{1+n_v}{2}}\Gamma(\frac{1+n_v}{2}), \quad \text{<Quantum-1>}
\\
\eta^{-\frac{1+n_v}{2}}\Gamma(\frac{1+n_v}{2}), \quad \text{<Real-Quantum-1>}
\end{cases}
\end{align}
where $\Gamma$ is the gamma function. The renormalized mass $\eta$ is fixed by $a^2m^2/2=0.1$ in \eqref{eq:eta} so that for in 1D spacetime,
\begin{align}
\eta &=
\begin{cases}
-0.9, \quad \text{<Quantum-1>}
\\
1.1, \quad \text{<Real-Quantum-1>}.
\end{cases}
\end{align}

\begin{figure}
    \centering
    \includegraphics[width=1.0\textwidth]{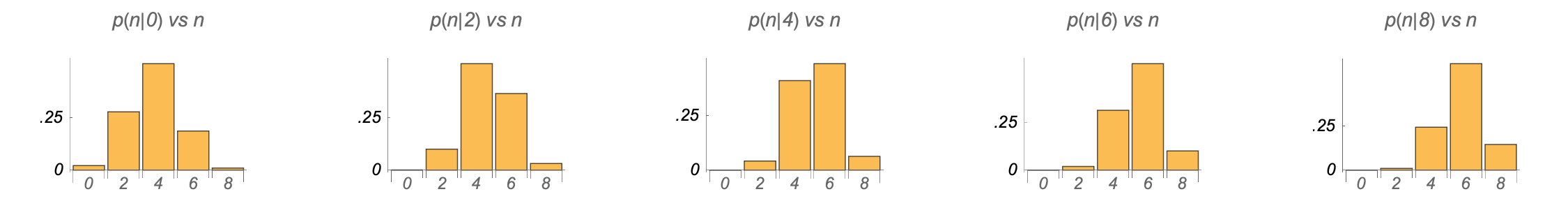}
    \caption{Test for the renormalization factor. Choice of the first value in \eqref{eq:evn}. This choice is not adopted.}
    \label{fig:renormalization1}
\end{figure}

\begin{figure}
    \centering
    \includegraphics[width=1.0\textwidth]{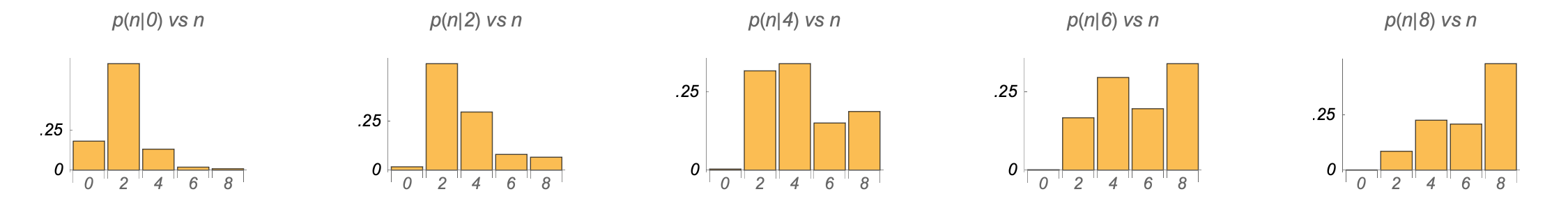}
    \caption{Test for the renormalization factor. Choice of the second value in \eqref{eq:evn}. This choice is not adopted.}
    \label{fig:renormalization2}
\end{figure}

\begin{figure}
    \centering
    \includegraphics[width=1.0\textwidth]{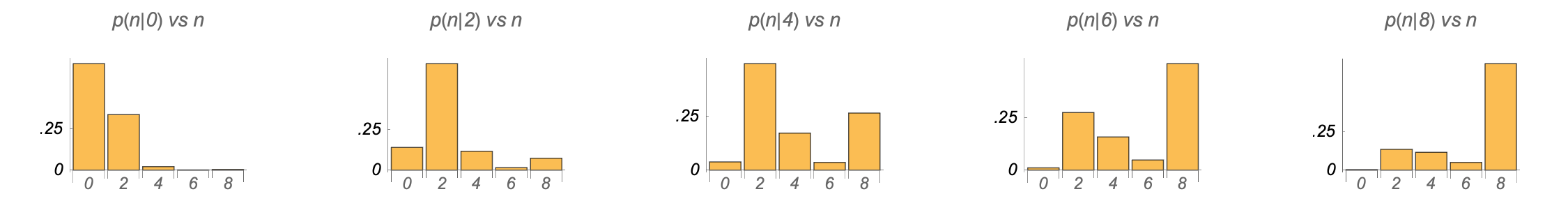}
    \caption{Test for the renormalization factor. Choice of the third value in \eqref{eq:evn}. This choice is not adopted.}
    \label{fig:renormalization3}
\end{figure}

\begin{figure}
    \centering
    \includegraphics[width=1.0\textwidth]{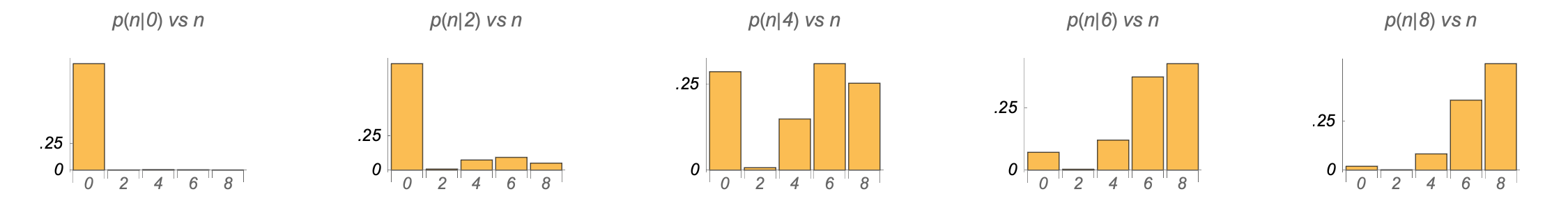}
    \caption{Test for the renormalization factor. Choice of the fourth value in \eqref{eq:evn}. This choice is not adopted.}
    \label{fig:renormalization4}
\end{figure}

\begin{figure}
    \centering
    \includegraphics[width=1.0\textwidth]{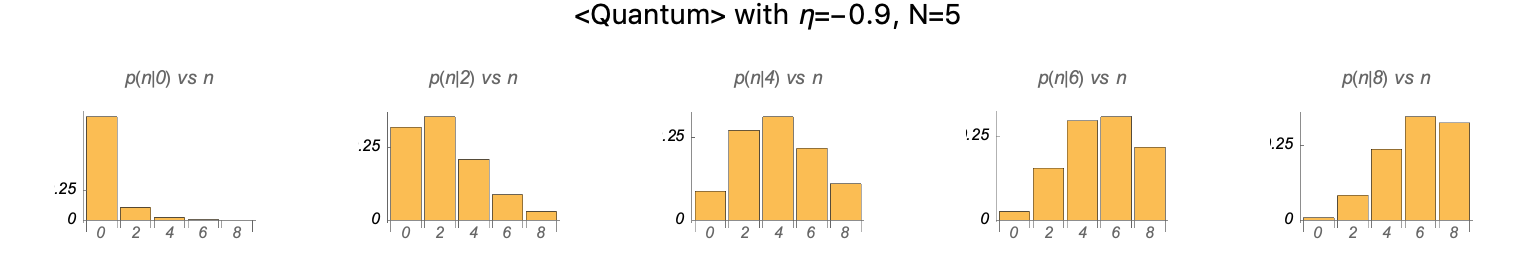}
    \caption{Test for the renormalization factor. Choice of the fifth value in \eqref{eq:evn}. This choice is adopted so the figure shows the results of $p(n|m)$ for $N=5$.}
    \label{fig:Quantum_No=5_eta=-0.9}
\end{figure}

Finally, a renormalization procedure is needed. To see this, note that we need to solve \eqref{eq:uMu} for $u$ by identifying the eigenvector of $M$ with eigenvalue $1$. However, plugging in the data $\eta=-0.9, N=5$ for instance and solving the eigensystem of $M$ numerically yields only the eigenvalues
\begin{align}\label{eq:evn}
&\{0.00132099,0.032359,0.292563,1.17831,2.14224\}\times (1+i).
\end{align}
In order to allow for eigenvalue $1$, we can renormalize by dividing out one of the above eigenvalues in the edge factor $E$ or the vertex factor $V$, since by the definition \eqref{eq:Mij} of $M$ all its elements is divided by the same factor whence $1$ becomes an eigenvalue. Yet which among \eqref{eq:evn} should one pick to divide out?
For this example of $N=5$, we compute $p(n|m)$ within <Quantum-1> for each of the  choices to obtain the results shown in \Cref{fig:renormalization1} to \Cref{fig:Quantum_No=5_eta=-0.9}. The last eigenvalue \eqref{eq:evn} with the largest modulus turns out to exhibit reasonable correlations so that for fixed $m$, $p(n|m)$ is peaked smoothly around $n=m$. The same test can be performed for other values of $N$, and we find as a rule of thumb that dividing out the eigenvalue with the largest modulus yields reasonable correlations. We will therefore divide by the largest eigenvalue for the renormalization.


\subsection{<Quantum-1> vs. <Real-Quantum-1>}

\begin{figure}
    \centering
    \includegraphics[width=1.0\textwidth]{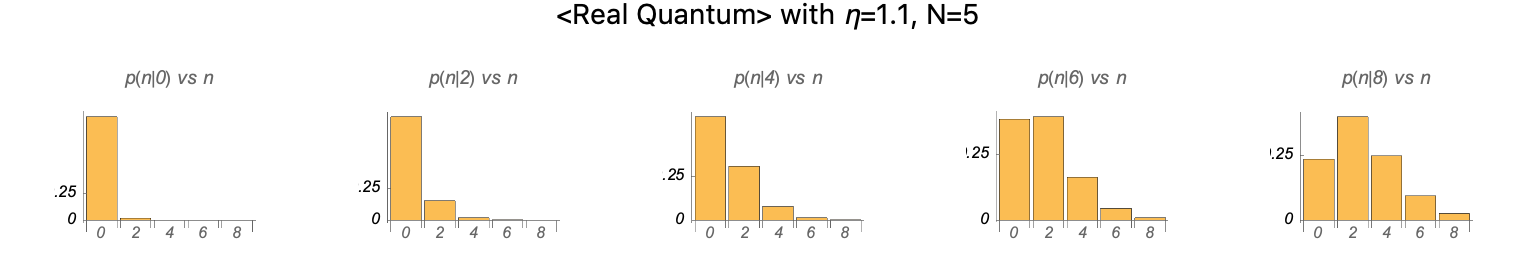}
    \caption{$p(n|m)$ for <Real-Quantum-1> with $N=5$.}
    \label{fig:Real-Quantum_No=5_eta=1.1}
\end{figure}

\begin{figure}
    \centering
    \includegraphics[width=0.5\textwidth]{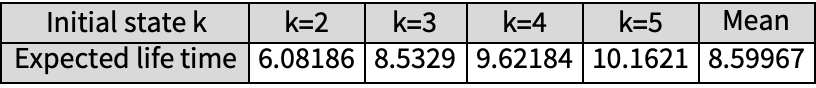}
    \caption{Life expectancy for <Quantum-1> with different initial particle numbers $m=2k-2$.}
    \label{fig:Quantum_No=5_eta=-0.9_life}
\end{figure}

\begin{figure}
    \centering
    \includegraphics[width=0.5\textwidth]{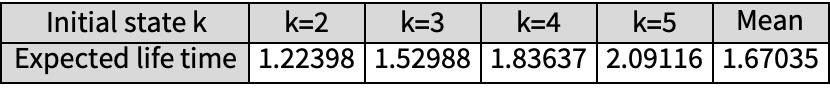}
    \caption{Life expectancy for <Real-Quantum-1> with different initial particle numbers $m=2k-2$.}
    \label{fig:Real-Quantum_No=5_eta=1.1_life}
\end{figure}

With the above setup, we can finally compute and compare the life expectancies of <Quantum-1> and <Real-Quantum-1>. For $N=5$, the probabilities $p(n|m)$ for <Quantum-1> and <Real-Quantum-1> are shown in \Cref{fig:Quantum_No=5_eta=-0.9} and \Cref{fig:Real-Quantum_No=5_eta=1.1}. From these we can numerically compute their life expectancies according to \eqref{eq:le} with the results shown in  \Cref{fig:Quantum_No=5_eta=-0.9_life} and \Cref{fig:Real-Quantum_No=5_eta=1.1_life}. It is seen that <Quantum-1> outlives <Real-Quantum-1>. This result is easily explained by comparing \Cref{fig:Quantum_No=5_eta=-0.9} and \Cref{fig:Real-Quantum_No=5_eta=1.1}. We see that <Quantum-1> whose path integral exponent is imaginary tends to stay at the same particle number as time passes, while <Real-Quantum-1>  whose path integral exponent is real tends to decay to lower particle numbers.


\begin{figure}
    \centering
    \includegraphics[width=1.0\textwidth]{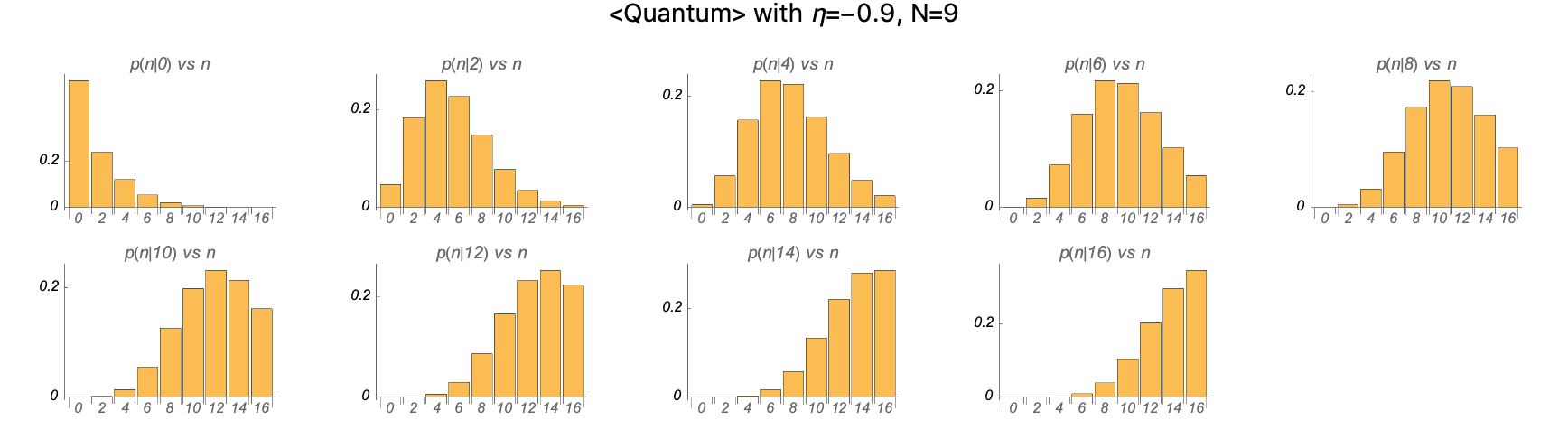}
    \caption{$p(n|m)$ for <Real-Quantum-1> with $N=9$.}
    \label{fig:Quantum_No=9_eta=-0.9}
\end{figure}

\begin{figure}
    \centering
    \includegraphics[width=1.0\textwidth]{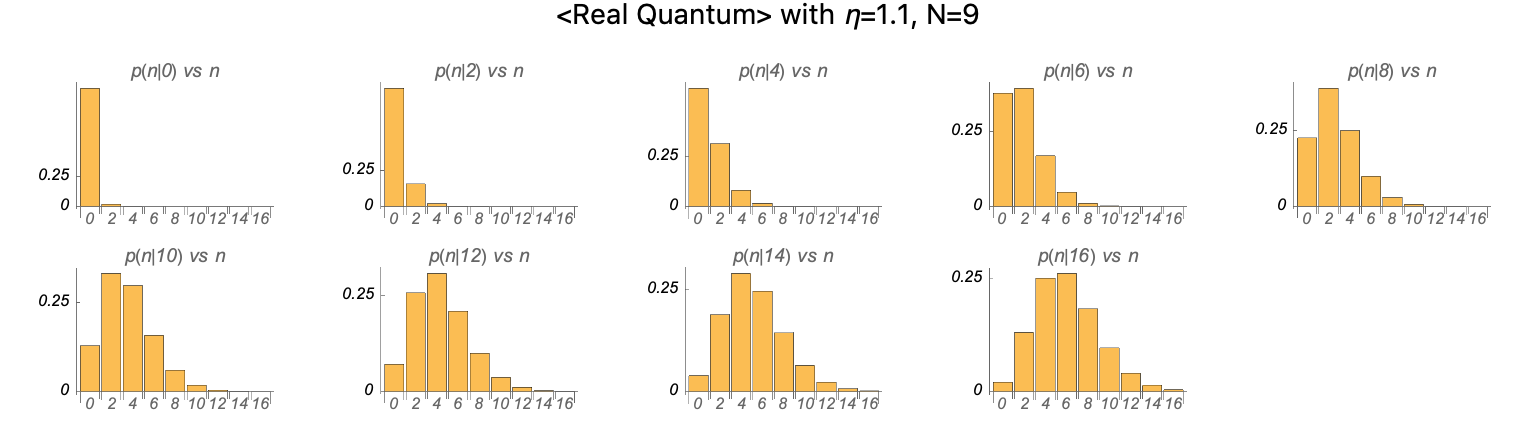}
    \caption{$p(n|m)$ for <Real-Quantum-1> with $N=9$.}
    \label{fig:Real-Quantum_No=9_eta=1.1}
\end{figure}

\begin{figure}
    \centering
    \includegraphics[width=0.9\textwidth]{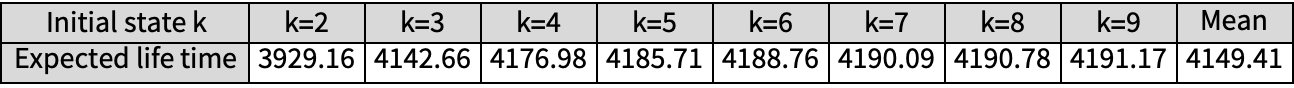}
    \caption{Life expectancy for <Quantum-1> with different initial particle numbers $m=2k-2$.}
    \label{fig:Quantum_No=9_eta=-0.9_life}
\end{figure}

\begin{figure}
    \centering
    \includegraphics[width=0.9\textwidth]{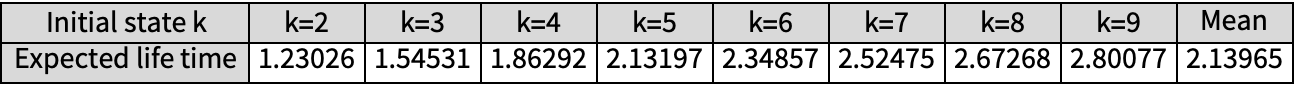}
    \caption{Life expectancy for <Real-Quantum-1> with different initial particle numbers $m=2k-2$.}
    \label{fig:Real-Quantum_No=9_eta=1.1_life}
\end{figure}

\begin{figure}
    \centering
    \includegraphics[width=1.0\textwidth]{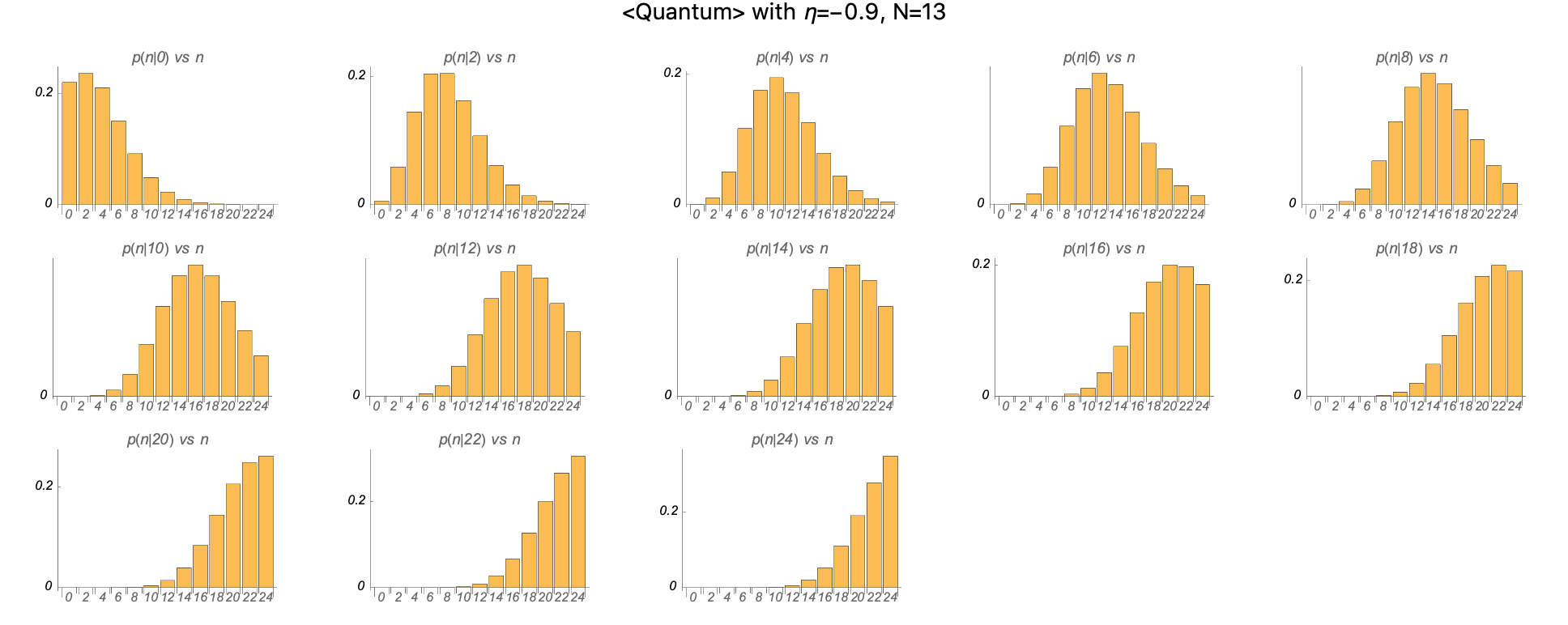}
    \caption{$p(n|m)$ for <Real-Quantum-1> with $N=13$.}
    \label{fig:Quantum_No=13_eta=-0.9}
\end{figure}

\begin{figure}
    \centering
    \includegraphics[width=1.0\textwidth]{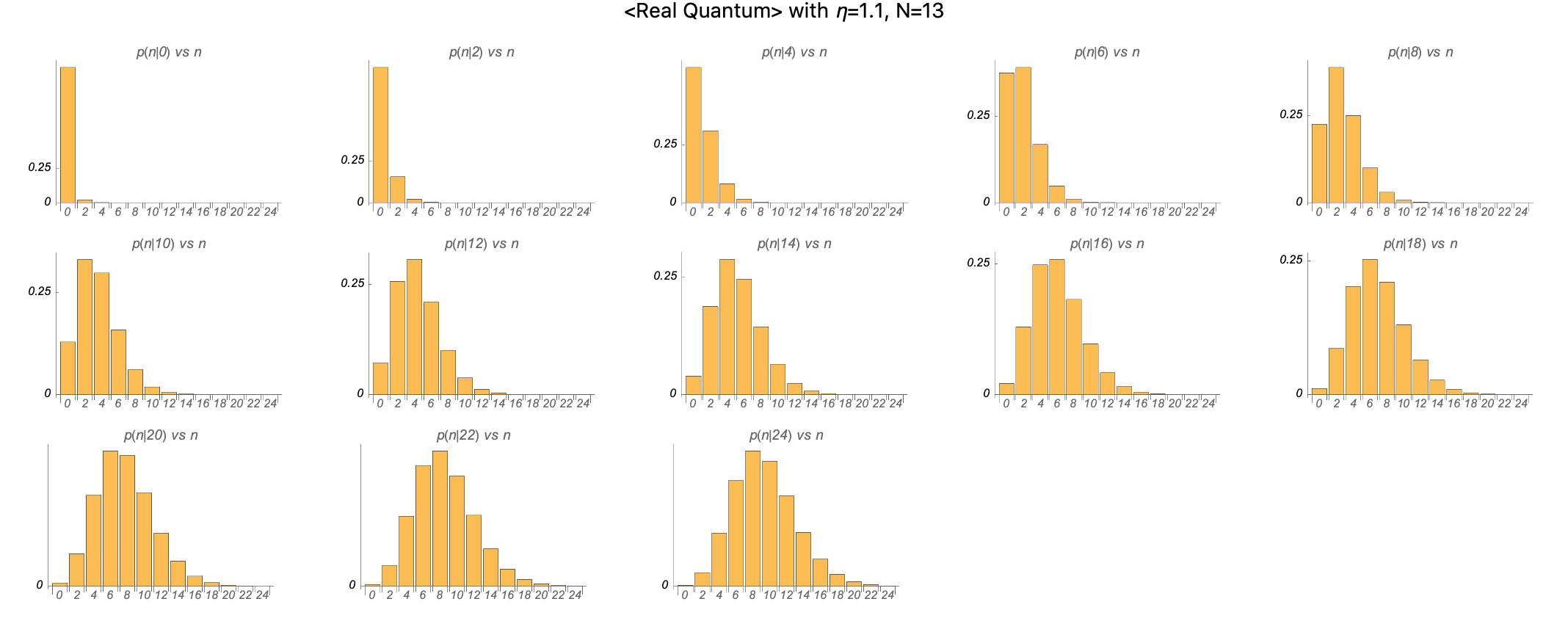}
    \caption{$p(n|m)$ for <Real-Quantum-1> with $N=13$.}
    \label{fig:Real-Quantum_No=13_eta=1.1}
\end{figure}

\begin{figure}
    \centering
    \includegraphics[width=1\textwidth]{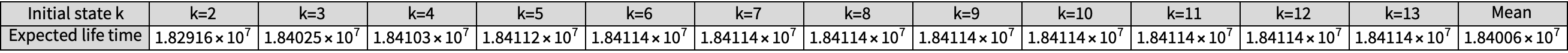}
    \caption{Life expectancy for <Quantum-1> with different initial particle numbers $m=2k-2$.}
    \label{fig:Quantum_No=13_eta=-0.9_life}
\end{figure}

\begin{figure}
    \centering
    \includegraphics[width=1\textwidth]{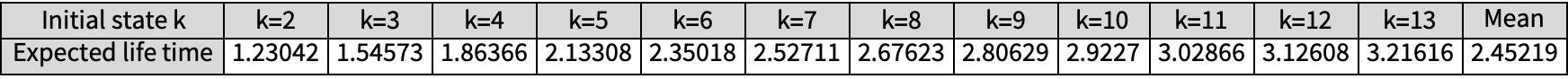}
    \caption{Life expectancy for <Real-Quantum-1> with different initial particle numbers $m=2k-2$.}
    \label{fig:Real-Quantum_No=13_eta=1.1_life}
\end{figure}

The same qualitative results hold for higher $N$ values such as $N=9$ and $N=13$ (\Cref{fig:Quantum_No=9_eta=-0.9} to \Cref{fig:Real-Quantum_No=13_eta=1.1_life}). Still <Quantum-1> tends to stay at the same particle number as time passes while <Real-Quantum-1> tends to decay quickly to lower particle numbers.



\subsection{<Quantum-1> with vs. without macroscopic superposition}

Next we compare <Quantum-1> with vs. without macroscopic superposition. In the literature there are multiple ways to define macroscopic superposition \cite{Frowis2018MacroscopicImplementations}. Here we refer to macroscopic superposition in a dynamical sense. From the previous results for $p(n|m)$ for <Quantum-1> it is seen that a state at a fixed particle number $m$ tends to evolve to states with particle numbers $n$ close to $m$. In other words, it would tend to take many time steps to change to a distant particle number. By macroscopic superposition, we mean a superposition of distant particle numbers. 

There have certainly been studies that attempt to show the impossibility of experiencing macroscopic superpositions. However, as far as I know these require quantum interpretational inputs, none of which is completely satisfactory \cite{Wallace2020OnProblem, Kent2010OneConfirmation}. One may naively think that even without quantum interpretational inputs, experiences of macroscopic superpositions can be ruled out just based on decoherence. This is false. As stressed for example by Schlosshauer, decoherence in itself does not solve the ``problem of definite outcomes'' \cite{Schlosshauer2003DecoherenceMechanics}. A reduced density matrix that is approximately diagonal does not imply the perception of an individual element. One still needs to supplement additional inputs to explain why and how single outcomes instead of macroscopic superpositions are perceived. In the words of Penrose ,
\begin{quote}
[...] there is nothing in the formalism of quantum mechanics that demands that a state of consciousness cannot involve the simultaneous perception of a live and a dead cat.
\end{quote}

In this work, we refrain from making any interpretation that excludes the possibility of experiences of macroscopic superpositions from the outset, and ask if there are evolutionary explanations.

\begin{figure}
    \centering
    \includegraphics[width=1\textwidth]{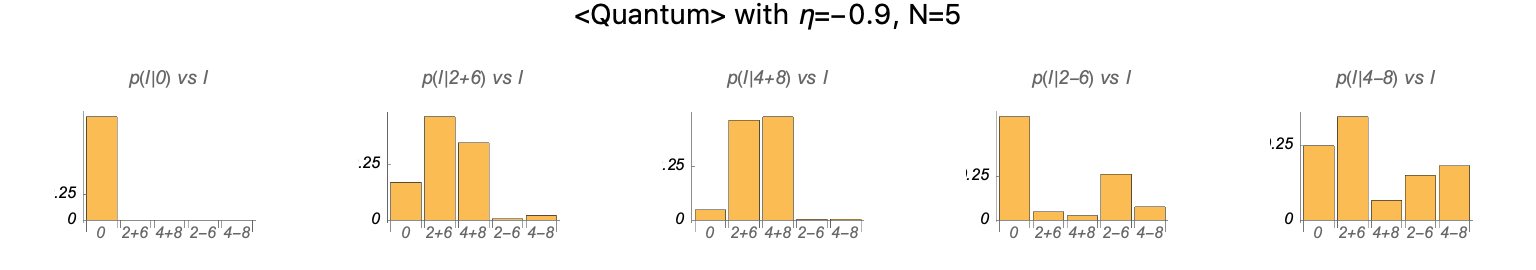}
    \caption{$p(l|l')$ for <Quantum-1> with macroscopic superpositions with $N=5$.}
    \label{fig:Quantum_No=5_eta=-0.9_SP}
\end{figure}

\begin{figure}
    \centering
    \includegraphics[width=0.5\textwidth]{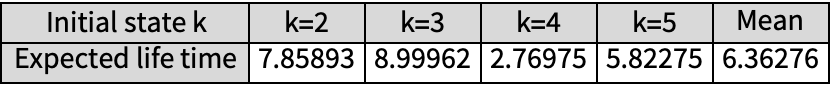}
    \caption{Life expectancy for <Quantum-1> with macroscopic superpositions with different initial particle numbers $m=2k-2$.}
    \label{fig:Quantum_No=5_eta=-0.9_SP_life}
\end{figure}

\begin{figure}
    \centering
    \includegraphics[width=1\textwidth]{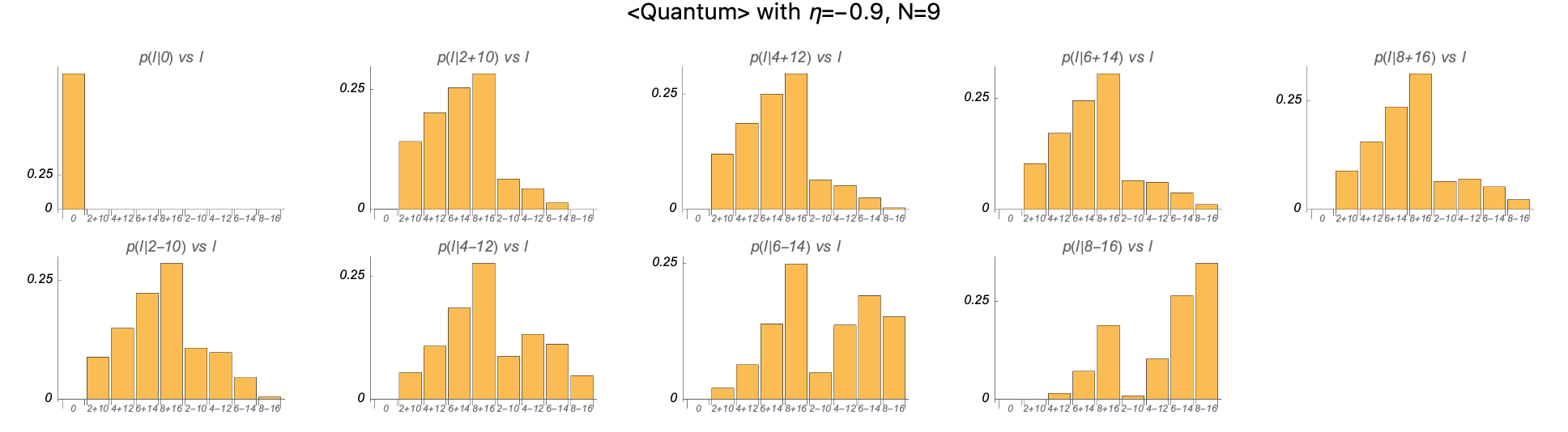}
    \caption{$p(l|l')$ for <Quantum-1> with macroscopic superpositions with $N=9$.}
    \label{fig:Quantum_No=9_eta=-0.9_SP}
\end{figure}

\begin{figure}
    \centering
    \includegraphics[width=0.9\textwidth]{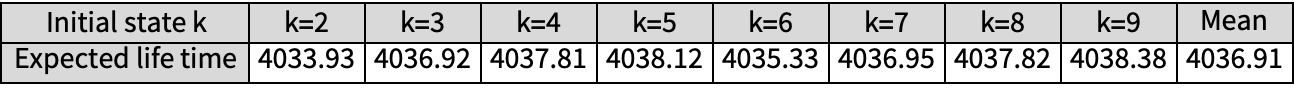}
    \caption{Life expectancy for <Quantum-1> with macroscopic superpositions with different initial particle numbers $m=2k-2$.}
    \label{fig:Quantum_No=9_eta=-0.9_SP_life}
\end{figure}

\begin{figure}
    \centering
    \includegraphics[width=1\textwidth]{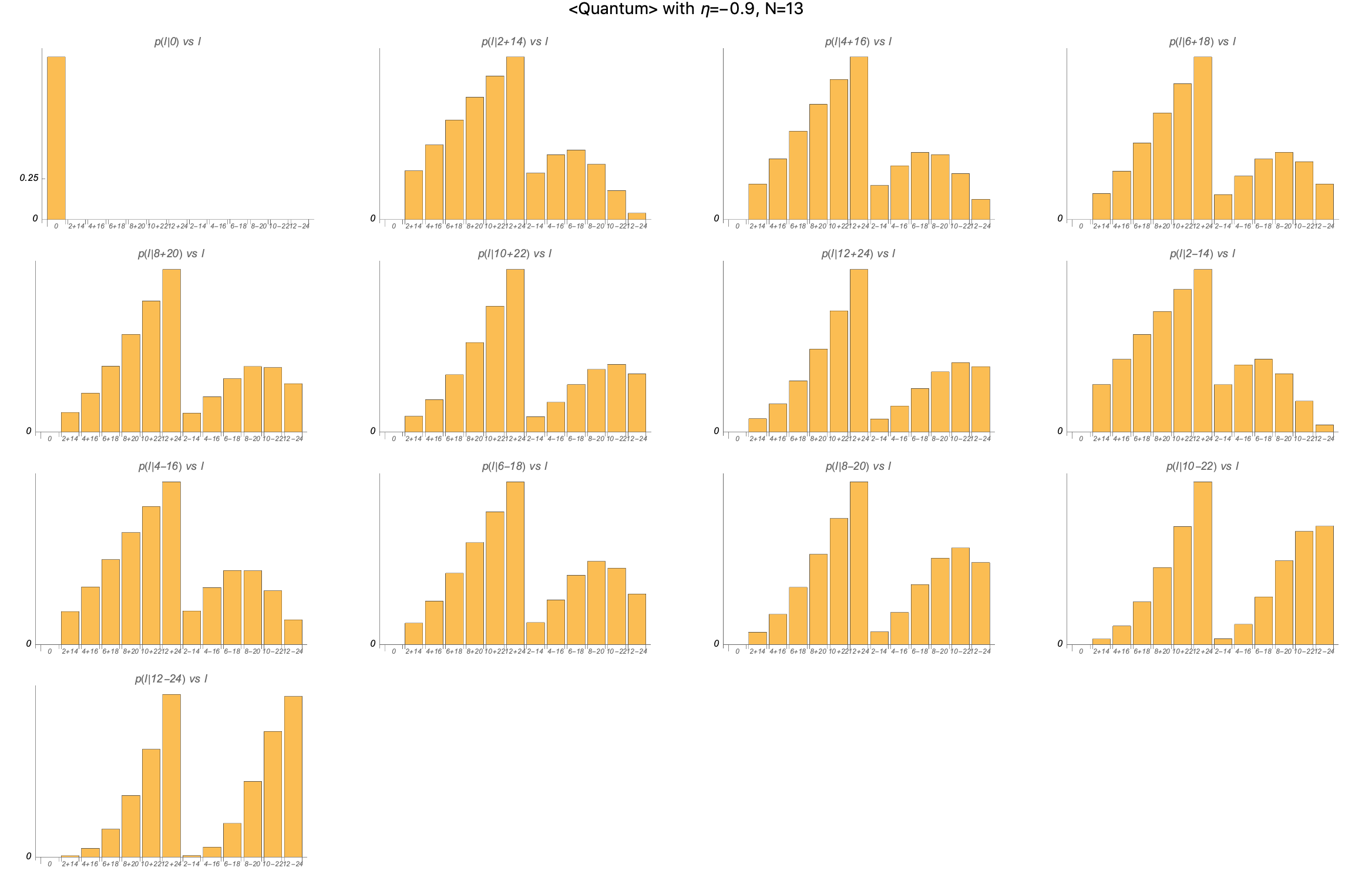}
    \caption{$p(l|l')$ for <Quantum-1> with macroscopic superpositions with $N=13$.}
    \label{fig:Quantum_No=13_eta=-0.9_SP}
\end{figure}

\begin{figure}
    \centering
    \includegraphics[width=1.0\textwidth]{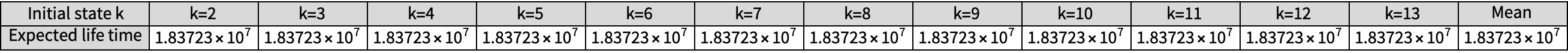}
    \caption{Life expectancy for <Quantum-1> with macroscopic superpositions with different initial particle numbers $m=2k-2$.}
    \label{fig:Quantum_No=13_eta=-0.9_SP_life}
\end{figure}

For the results presented in \Cref{fig:Quantum_No=5_eta=-0.9_SP} to \Cref{fig:Quantum_No=13_eta=-0.9_SP_life}, we consider states of the form 
\begin{align}
l=l_1\pm l_2,
\end{align}
which correspond to $\ket{l}=\frac{1}{\sqrt{2}}(\ket{l_1}+\ket{l_2})$ in the usual Hilbert space notation. Here the particle numbers are taken to be $l_1=2,\cdots,N-1$ and $l_2=l_1+N-1$ for each $l_1$.

By comparing the life expectancies it is seen that <Quantum-1> without macroscopic superposition outlives <Quantum-1> with macroscopic superposition, although by much smaller margins than over <Real-Quantum-1>. The explanation can be found by inspecting the probabilities $p(l|l')$ for <Quantum-1> with macroscopic superposition. In comparison to <Quantum-1> without macroscopic superposition, the dynamics tends to push the state further away from its current state, which increases the chances for a state to evolve to the death state even when the original state is far away from the death state. 

An informal intuition is that ordinarily an average being is born in a state far from death, and evolves gradually towards death as it ages. On the other hand when allowed to experience macroscopic superpositions of, for instance, living at a young age and living at a middle age near the beginning of a being's life, the life expectancy is shortened because the component at a middle age makes the being accelerate towards death.




\subsection{Towards more realistic models}\label{sec:bsm}

\begin{figure}
    \centering
    \includegraphics[width=0.6\textwidth]{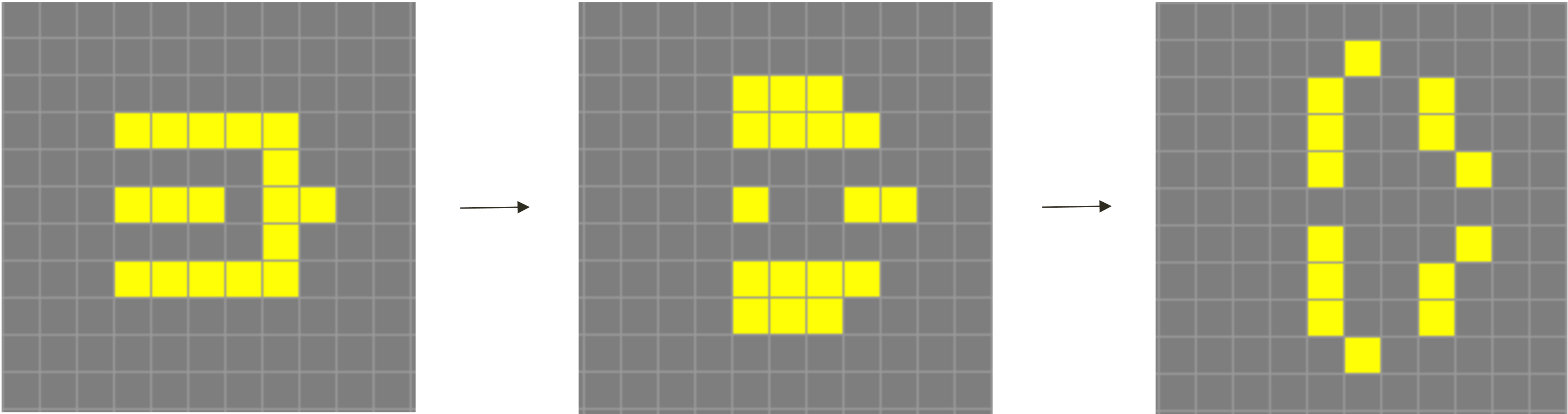}
    \caption{A process in Conway's Game of Life. Image produced with the website: https://playgameoflife.com/.}
    \label{fig:gol}
\end{figure} 


The above toy model is certainly not realistic in modelling actual beings such as human beings with far more sophisticated brain configurations. However they provide easy-to-work with models for us to harness and test ideas that may apply in realistic models. In this sense, the model may be compared to Convey's Game of Life \cite{Gardner1970Mathematicallife} (\Cref{fig:gol}) that describes a simplified lower-dimensional world where matter can lump together to survive or annihilate to die, which have certainly been fruitful in suggesting ideas for more sophisticated studies on, for instance, the topics of complexity and emergence for living organisms.

It is reasonable to try and situate the present models of life and death even closer to a realistic theory of our universe, especially given that the present model is based on scalar field theories, which are quite close to the realistic Standard Model. Here are some possible steps to take:
\begin{itemize}
\item Generalize the potentials in the actions.
\item Introduce one more degree of freedom to consider a complex scalar field, which shows up for the Higgs field of the Standard Model.
\item Introduce other fields of the Standard Model.
\item Path integrate over spacetime configurations for quantum gravity.
\item Take the continuum limit of the lattice theories.
\end{itemize}
In these developments, it could be helpful to adopt the particle (or particle-string) representations \cite{Gattringer2013SpectroscopyGas, Gattringer2013LatticeField, Gattringer2016ApproachesTheory, Jia2022WhatModel} of the field theories to obtain an integer basis description, which we did above for the real scalar field theory. 

In working with any of these more realistic theories, one could relax some simplifying assumptions made above for the present model:
\begin{itemize}
\item Generalize to higher dimensions of spacetime.
\item Consider bounded regions of spacetime and non-trivial boundary conditions.
\item Push up the cutoff at $N$.
\item Allow non-adjacent edges for sequential candidate experiences.
\item Model life and death states with more details.
\item Consider further evolutionary fitness functions in addition to life expectancy.
\end{itemize}
Finally, one could bring in additional modes of experience for evolutionary considerations, or even \textit{design} modes of experience to increase certain evolutionary fitness functions and investigate if these are realizable.


\section{Discussions}\label{sec:d}

In discussions of quantum physics, it is usually assumed that all beings experience a superposed world in the same mode for always. On the other hand, the quantum measurement problem has also been with us quantum physicists for always.

The problem may lie in the assumption. It could be that the universal laws of physics do not imply the Born rule, and that some beings (e.g., a futuristic intelligent quantum computer with consciousness) have experiences conforming to other rules. 

In this case attempts to extract explanations for human beings' experiences directly from the universal laws of physics are doomed to fail. Because the universal laws are also compatible with other modes of experience, the explanation can only be based on additional inputs.

In this scenario, one mode of experience is to be understood in the background of a variety of modes of experience. This is reminiscent of evolutionary biology, where the current form of a species is to be understood in the background a variety of forms that could be taken.

The results of \Cref{sec:ec} show that even though the ordinary quantum mode of experience -- <Quantum-1> without macroscopic superposition -- may not apply to all experiential beings, it may be preferred by natural selection. In this scenario, the evolutionary history of life on Earth (and elsewhere) is still more different than what is taught in contemporary biology with a largely classical worldview. There could be yet other living forms with alternative modes of experience that evolution probed in a superposed world.

Certainly the idea that there are alternative modes of experience is speculative. Yet I hope the results presented in this work illustrate some interesting prospects to be investigated further. 

I conclude with a discussion on some broader topics relevant for interpretations of quantum theory. 

\subsection{Ontology}\label{sec:o}

At the dawn of ancient Greek metaphysics, Parmenides delivered a startling view on the ontology of the universe \cite{Coxon2009TheParmenides}:
\begin{quote}
[...] that Being is ungenerated and imperishable, entire, unique, unmoved and perfect; it never was nor will be, since it is now all together, one, indivisible.
\end{quote}
This is in apparent conflict with ordinary perceptions of a multitude of things changing. Parmenides attributes this to the illusory appearances that we mortals gather:
\begin{quote}
Thus, I say, according to belief these things originated and now are and in later times hereafter, having received their sustenance, will end. On them men bestowed a name to give its mark to each. 
[...]
\end{quote}

The superposed world of \Cref{sec:sw}, and the modes of experiences of \Cref{sec:moe} which do not encompass all that exists in the superposed world (\Cref{sec:ennee}) are certainly reminiscent of Parmenides. 

\begin{figure}
    \centering
    \includegraphics[width=0.6\textwidth]{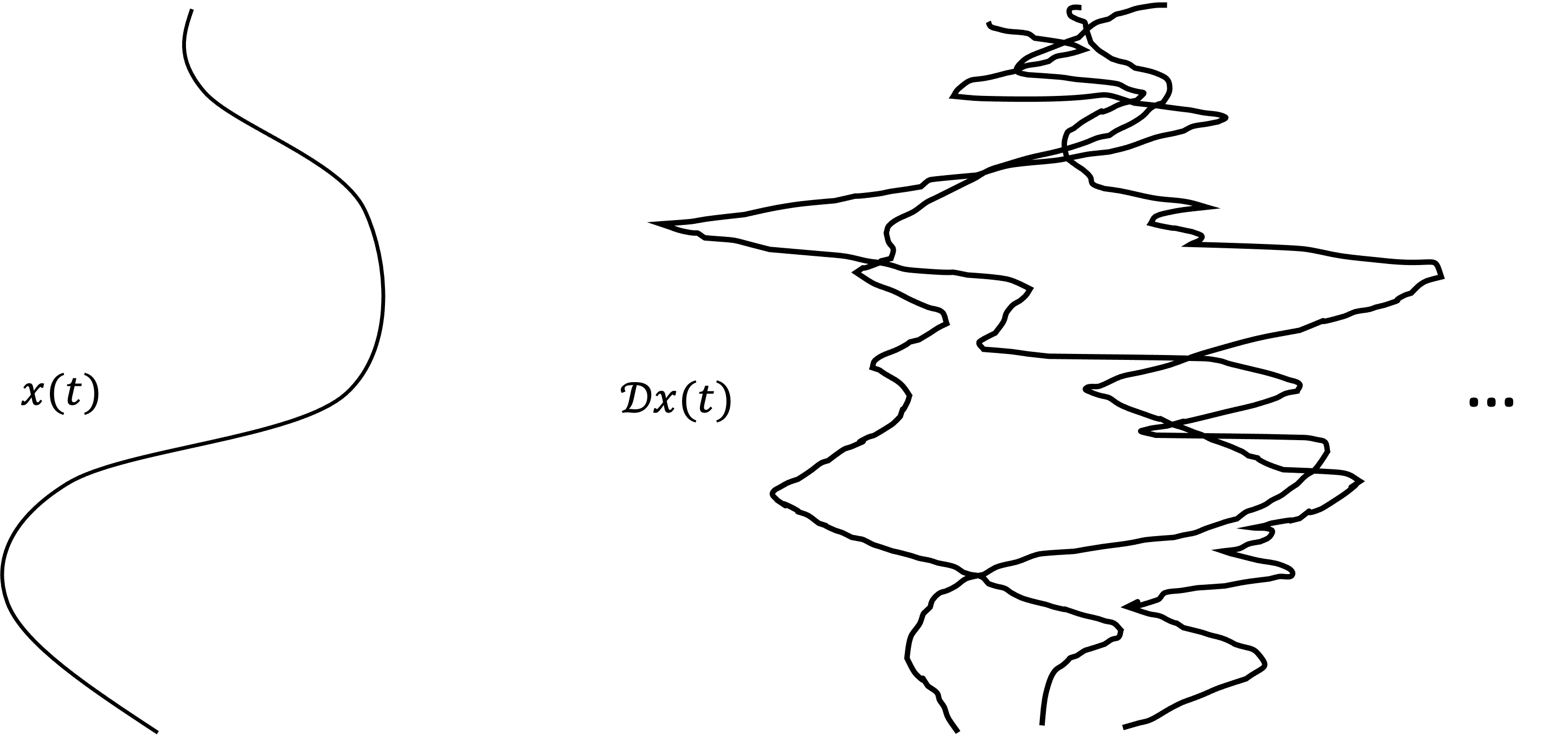}
    \caption{Left: a Some Ontology consisting of one particular particle history. Right: an All Ontology consisting of all particle histories.}
    \label{fig:sva}
\end{figure} 

If one is to fix an ontology for the superposed world of \Cref{sec:sw} and characterize its main feature, the obvious thing to note is that the superposed world includes \textit{all} physical configurations in the ontology.

We could accordingly distinguish two types of ontologies (\Cref{fig:sva}). In a universe with a \textbf{Some Ontology}, \textit{some} physical configuration(s) exist out of all the possibilities. This could be one particle or field history in classical mechanics, or one Hilbert space state evolution history out of many other possibilities in quantum theory. 
In contrast, in a universe with an \textbf{All Ontology}, \textit{all} possible physical configurations exist. 
For example, this holds for ontology of \eqref{eq:aos} which includes all physical configurations as path integral configurations. 

Metaphysically, a Some Ontology faces the tough question: ``Why does this/these exist out of all the possibilities?''. This question does not arise for an All Ontology, for all possibilities exist.

\subsection{Probability as propensity}\label{sec:ppa}

The conscious experience for human beings seems to exhibit a tendency for a coherent and unambiguous experience at the cost of hiding much sensory input \cite{Dehaene2014ConsciousnessThoughts}. For instance, in the phenomenon of binocular rivalry discussed in \Cref{sec:mle}, human beings are presented at each moment with a coherent and unambiguous conscious visual experience at the cost of hiding half of what is seen. 

Yet the perceived view also switches stochastically between the two images as time moves on. The ``sampling hypothesis'' is one idea to explain this. It holds that in the presence of perceptive ambiguities, that the mind draws samples among different possible perceptions given the sensory inputs according to some likelihood probability. In the version of the hypothesis of \cite{Icard2016SubjectivePropensity}, the probabilities are objective propensities for actualizing the different sampling possibilities.

There is certainly no shortage of perceptive ambiguities in a superposed world with an All Ontology. It could be that the probabilistic aspect of the experiences for humans and others has a similar explanation. The superposed world supplies multiple possibilities for experiences that result in ambiguities. When coherence is required for conscious experiences, the ambiguities are resolved by probabilistic sampling. Here the probabilities are neither frequencies nor beliefs, but propensities for actualizing the different sampling possibilities.

\subsection{On Theories of Everything}\label{sec:toe}

Don Page judiciously points out that a complete physical theory of a quantum universe have to include at least the following elements \cite{PageBornAgain}:
\begin{enumerate}
\item Kinematic variables
\item Dynamical laws
\item Boundary conditions
\item Specification of what has probabilities
\item Probability rules (analogue of Born’s rule)
\item Specification of what the probabilities mean
\end{enumerate}
Traditionally, only items 1 and 2 are considered in ``Theories of Everything'' where postulates are made on the fundamental kinematic variables and their dynamical laws. In quantum cosmology, postulates for the boundary condition of the universe such as the Hartle-Hawing boundary condition \cite{Hartle1983WaveUniverse} are investigated.

These are not enough in view of items 4 to 6. For items 4 and 6, \Cref{sec:ppa} suggest that probabilities could arise when experiential beings encounter experiential ambiguities. The probabilities could represent propensities for some subjective experience to actualize among the ambiguous possibilities. For item 5, the main message of this work is that we may need to specify different probability rules for different beings with distinct modes of experience.

This calls for a revised view on item 2, the dynamical laws. Traditionally, the probabilistic rules of \eqref{eq:quantum1} and \eqref{eq:quantum2} are viewed as part of the universal dynamical laws of Nature, while here they are categorized under item 5 as species-specific probability rules for experiences. In one way of understanding, the proposal made here is to consider the possibility that some previously believed universal dynamical laws are only species-specific.

\subsection{GPT and GET}\label{sec:fgpt}

General Probabilistic Theories (GPT) \cite{Popescu1994QuantumAxiom, HardyQuantumAxioms, barrett_information_2007} have been much studied in quantum foundations. They are ``general'' because in that framework we consider probabilistic rules for observations that go beyond that of quantum theory. The probabilistic rules in the modes of experience considered here also go beyond that of quantum theory, and in this sense one may refer to the present framework as a framework for General Experience Theories (GET).

A main difference is that in GPTs, the central object is 
\begin{align}
p(a,b,c,\cdots|x,y,z,\cdots),
\end{align}
the conditional probabilities for observing outcomes $a,b,c,\cdots$ given some choices of measurement settings $x,y,z,\cdots$. Here the choices and observations can be made by multiple agents, and $p(a,b,c,\cdots|x,y,z,\cdots)$ is part of a ``third person'' description. In contrast, the ``first person'' description of GET always refers to the conditional probabilities for an individual being's (candidate) experiences. Adopting a first person description may clarify studies of the Wigner's friend setting, which  received a renewed interest in recent years \cite{Brukner2017OnProblem, Brukner2018AFacts, Frauchiger2018QuantumItself} and calls for an unambiguous specification of individual beings' observations or experiences.

As another difference, in GPT we often think of universes governed entirely by \textit{one} probability rule, e.g., that of real Hilbert space quantum theory. In contrast, in GET we can consider universes where \textit{multiple} different modes of experience apply to different beings or the same being at different stages living in the \textit{same} universe. This enables evolutionary considerations as demonstrated in \Cref{sec:ec}.

Moreover, as a practical difference, in the considerations above for GET we fixed the physical configurations as path integral configurations, which suggests an explicit ontology (\Cref{sec:o}). In contrast, in GPTs one usually proceeds without fixing the physical configurations and remains ambiguous about the ontology, in the sense of not specifying if the matter in the world is made of field, particle etc. This is a practical difference because presumably one could also choose to fix the physical configurations in GPT, or remain ambiguous about the physical configurations in GET. 

In the evolutionary considerations of \Cref{sec:ec} the explicit physical configurations were needed in order to obtain quantitative results about life expectancy, which highlights what fixing the physical configurations allows one to explore.

On the other hand, GPTs are particularly helpful for deriving general structural properties of theories independent of the ontology. A similar study about general structural properties of modes of experiences in GET may also yield insights on experiences in a superposed world.

\subsection{Towards an interpretation of quantum theory}\label{sec:taqt}

The above discussions bring out some suggestive ideas towards an interpretation of quantum theory.
\begin{itemize}
\item \textbf{All Ontology.} What is the ontology? The basic ontology of the world is an All Ontology in the sense of \Cref{sec:o}. All possible physical configurations as characterized by some path integral exist in superposition.
\item \textbf{Propensity probabilities.} How do probabilities arise? Probabilities arise neither in a Frequentist nor Bayesian fashion. As discussed in \Cref{sec:ppa}, probabilities capture the objective propensities for certain subjective experiences to actualize, when alternatives are presented by the superposed world.
\item \textbf{Limited applicability.} Quantum theory does not apply universally to all experiences. Instead, experiences can conform to alternative probability rules like those exemplified in \Cref{sec:ex}. 
\item \textbf{Evolution.} Experiences conforming to the quantum mode or other modes may be explained by its evolutionary advantages conditioned on the environment.
\end{itemize}

These ideas are preliminary, and will likely need to be updated. One substantial task for theory development is understand the process of mutation for modes of experience towards a dynamical description for the evolution. See, for example, Kent \cite{Kent2016QuantaQualia} for some potentially relevant ideas. In addition, Smolin advanced the view that the dynamical laws of physics could evolve \cite{Smolin2013TimeUniverse}. As discussed in \Cref{sec:toe}, in the present framework the rules for experiences subsumes some roles of dynamical laws. Evolving modes of experiences may provide a new perspective to realize evolving laws.

\section*{Acknowledgement}

I am very grateful to Lucien Hardy and Matthew Fox for discussions on the topics under study, to Lee Smolin for conversations on related broader topics, and to Lucien Hardy and Achim Kempf for long-term encouragement and support. Research at Perimeter Institute is supported in part by the Government of Canada through the Department of Innovation, Science and Economic Development Canada and by the Province of Ontario through the Ministry of Economic Development, Job Creation and Trade. 

\appendix

\section{Scalar field path integrals}\label{sec:sfpi}

\subsection{Particle representation for scalar field theory}

Consider a real scalar field theory in Minkowski spacetime with the Lagrangian density
\begin{align}
    \mathcal{L}=-\frac{1}{2}\partial^\nu \phi \partial_\nu \phi-\frac{1}{2} m^2\phi^2(x) - V(\phi)
\end{align}
with a general potential $V$. Here the metric signature convention is
\begin{align}
(-,+,+,\cdots).
\end{align}
We want to define the path integral non-perturbatively, and the standard procedure is through a lattice \cite{Peskin1995}. Let there be a $D$-dimensional hypercubic lattice with spacing $a$ in both time and space directions. Rewriting derivatives as differences, we obtain the lattice action
\begin{align}
S=&a^D \sum_x [ -\frac{1}{2} \sum_{\nu=1}^D g^{\nu\nu}(\frac{\phi_{x+\nu}-\phi_x}{a})^2 - \frac{1}{2} m^2 \phi_x^2 - V(\phi_x)]
\\
=& \sum_x [ \sum_{\nu=1}^D g^{\nu\nu} \tilde{\phi}_{x+\nu} \tilde{\phi}_{x} - \eta \tilde{\phi}_x^2  - \tilde{V}(\tilde{\phi}_x)].\label{eq:rsal3}
\end{align}
Here $g^{\nu\nu}$ is the Minkowski metric, $x$ refers to lattice vertices, and $x\pm \nu$ refers to the vertex one unit in the positive or negative $\nu$-th direction away from $x$.
In the last line, we introduced 
\begin{align}
\tilde{\phi}_x &= a^{\frac{D-2}{2}}\phi_x,
\\
\eta &=a^2 m^2/2 + D-2,
\\
\tilde{V}(\tilde{\phi}_x)&=a^D V(\phi_{x}).
\end{align}
The tilde symbols are omitted in the following for simplicity. The path integral partition function is given by
\begin{align}\label{eq:pfrsf}
    Z=\int D\phi e^{iS}.
\end{align}
The lattice spacing limit $a\rightarrow 0$ needs to be taken if one wants to obtain results for continuum spacetime.

\subsection*{Particle representation}

For the following studies it is convenient to use a particle representation for the above field path integral \cite{Gattringer2013SpectroscopyGas, Gattringer2013LatticeField, Gattringer2016ApproachesTheory, Jia2022WhatModel}. Let $S_{1}$ be the first term of (\ref{eq:rsal3}). In the notation $\prod_{x,\nu}:= \prod_x \prod_{\nu=1}^D$ and $\sum_{n}:=\prod_{x,\nu} \sum_{n_{x,\nu}=0}^\infty$,
\begin{align}
e^{iS_{1}}=&\prod_{x,\nu} \exp{ i g^{\nu\nu} \phi_{x+\nu} \phi_{x} }
=\sum_{n}\prod_{x,\nu} \frac{( i g^{\nu\nu}\phi_{x+\nu} \phi_{x})^{n_{x,\nu}} }{n_{x,\nu}!} 
\\
=&\sum_{n}
(\prod_{x,\nu}\frac{( i g^{\nu\nu})^{n_{x,\nu}}}{n_{x,\nu}! })
(\prod_{x} \phi_x^{\sum_{\nu=1}^D(n_{x,\nu}+n_{x-\nu,\nu})}),
\\
Z=\int D\phi ~ e^{iS}=& \mathcal{N} \sum_{n} (\prod_{x,\nu} \frac{( i g^{\nu\nu})^{n_{x,\nu}}}{n_{x,\nu}!}) (\prod_{x} \int_{-\infty}^\infty d\phi_x ~\phi_x^{n_x} e^{-i \eta \phi_x^2-i V(\phi_x)}) \label{eq:rsfz}
\\
=& \mathcal{N} \sum_{n} (\prod_{x,\nu} \frac{( i g^{\nu\nu})^{n_{x,\nu}}}{n_{x,\nu}!}) (\prod_{x} f(n_x)),
\label{eq:rsfz1}
\end{align}
where $\mathcal{N}$ is a constant, $f$ stands for the last integral of \eqref{eq:rsfz}, and $n_x:=\sum_{\nu=\pm 1}^{\pm D} n_{x,\nu}$. Here $Z$ is defined without the lattice spacing limit $a\rightarrow 0$ as for the toy model studied here we are contend with lattice results for simplicity.

\begin{figure}
    \centering
    \includegraphics[width=1.0\textwidth]{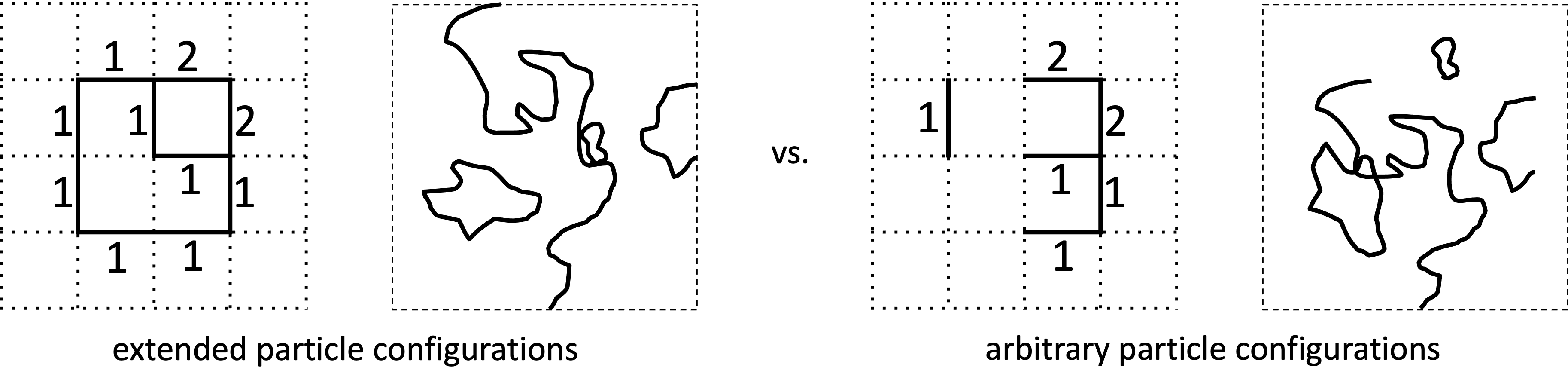}
    \caption{Left: extended particle configurations for a $Z_2$-symmetric theory. Right: arbitrary particle configurations for a general theory.}
    \label{fig:pwl2}
\end{figure} 

In the first line the exponential for the kinetic coupling term is Taylor expanded, which introduces an integer variable $n_{x,\nu}$ on the edge $x,\nu$ connecting $x$ and $x+\nu$. This mathematically trivial step is physically profound, as the basic entity is changed from fields to particles. Now the path integral sum \eqref{eq:rsfz1} is over $n$-configurations assigning non-negative integers $n_{x,\nu}$ to the lattice edges. The integer $n_{x,\nu}$ represents the number of particles passing the edge $x,\nu$, and $n_x$ as the total number of particle line segments passing $x$. 

In this work I consider scalar field theories exhibiting a global $Z_2$ symmetry so that $V(r)=V(-r)$. This implies that the particle lines keep extending instead of popping in or out of existence at the vertices. To see this, note that
\begin{align}
f(n_x)=& \int_{-\infty}^\infty d\phi_x ~\phi_x^{n_x} e^{-i \eta \phi_x^2-i V(\phi_x)}
\\
=& \int_0^\infty d r_x ~ r_x^{n_x}e^{-i\eta r_x^2}[e^{-iV(r_x)}+(-1)^{n_x}e^{-iV(-r_x)}]
\\
=& \int_0^\infty d r_x ~ r_x^{n_x}e^{-i\eta r_x^2}[(1+(-1)^{n_x})e^{-iV(r_x)}]
\\
=& 2 \delta_2(n_x)\int_0^\infty d r_x ~ r_x^{n_x}e^{-i\eta r_x^2-iV(r_x)},
\\
Z=& \mathcal{N}\sum_{n}(\prod_{x,\nu} \frac{( i g^{\nu\nu})^{n_{x,\nu}}}{n_{x,\nu}!}) (\prod_{x} 2 \delta_2(n_x)\int_0^\infty d r_x ~ r_x^{n_x}e^{-i\eta r_x^2-iV(r_x)}).
\end{align}
In the second line the $\phi_x$ integral is separated into the positive and negative parts and recombined. In the third line the $Z_2$ symmetry is used. In the fourth line $\delta_2(x)$ is the mod 2 Kronecker delta function, which arises because $1+(-1)^{n_x}=0$ for odd $n_x$. The presence of $\delta_2(n_x)$ at all vertices in the last line implies that the number of particles crossing any vertex is always even, which admits the interpretation that the particle lines keep extending and are never ending. Therefore the particle configurations summed over are as on the left of \Cref{fig:pwl2}.

Finally, we rewrite the path integral for a $Z_2$-symmetric theory in a tidier form
\begin{align}
Z=& \mathcal{N}\sum_{n\text{ extended}} \prod_{e} E_e(n_e) \prod_{v} V_v(n_v),
\end{align}
where $\sum_{n\text{ extended}}$ sums over extended particle configurations, $e$ relabels the lattice edges $(x,\nu)$, $v$ relabels the lattice vertices $x$, and
\begin{align}
E_e(n_e)=&\frac{( i g^{e})^{n_{e}}}{n_{e}!},
\\
V_v(n_v)=&2\int_0^\infty d r ~ r^{n_v}e^{-i\eta r^2-iV(r)}
\end{align}
are the edge and vertex amplitudes.

\subsection{Real-valued path integrals}

In addition to $Z=\int D\phi e^{iS}$ with complex amplitude, in this work I also consider path integrals with real amplitudes of the form
\begin{align}\label{eq:pfrsf2}
    Z=\int D\phi e^{-S}.
\end{align}
For a real scalar field theory consider the Lagrangian density
\begin{align}
    \mathcal{L}=\frac{1}{2}\partial^\nu \phi \partial_\nu \phi+\frac{1}{2} m^2\phi^2(x) + V(\phi)
\end{align}
with a general potential $V$. To define the theory non-perturbatively we introduce a $D$-dimensional hypercubic lattice with spacing $a$ in both time and space directions. The lattice action is
\begin{align}
S=&a^D \sum_x [\frac{1}{2} \sum_{\nu=1}^D g^{\nu\nu}(\frac{\phi_{x+\nu}-\phi_x}{a})^2 + \frac{1}{2} m^2 \phi_x^2 + V(\phi_x)]
\\
=& \sum_x [-\sum_{\nu=1}^D g^{\nu\nu} \tilde{\phi}_{x+\nu} \tilde{\phi}_{x} + \eta \tilde{\phi}_x^2  + \tilde{V}(\tilde{\phi}_x)].\label{eq:rsal23}
\end{align}
In the last line, I introduced
\begin{align}
\tilde{\phi}_x&= a^{\frac{D-2}{2}}\phi_x
\\
\eta&=a^2 m^2/2 + \sum_{\nu=1}^D g^{\nu\nu}
\\
\tilde{V}(\tilde{\phi}_x)&=a^D V(\phi_{x}).
\end{align} 
The tilde symbols are omitted in the following for simplicity. The path integral partition function is given by \eqref{eq:pfrsf2}. The lattice spacing limit $a\rightarrow 0$ needs to be taken if one wants to obtain results for continuum spacetime.

The metric is taken to be
\begin{align}\label{eq:mrvpi}
g^{\mu\nu}=\delta^{\mu\nu}
\end{align}
so that the partition function agrees with that of Euclidean QFT. In Euclidean QFT, an inverse Wick rotation is required to map the Euclidean theory back to Minkowski spacetime. In contrast, here the theory with the partition function \eqref{eq:pfrsf2} is defined directly in Minkowski spacetime. No Wick rotation or inverse Wick rotation is ever needed. This distinguishes the present Minkowski spacetime theory with real-valued path integrals from Euclidean QFT in principle, although in practice results from Euclidean QFT can be used for the present theory without the need to ever perform inverse Wick rotations.


\subsection*{Particle representation}

The reformulation into the particle representation proceeds similarly as above. Let $S_{1}$ be the first term of (\ref{eq:rsal23}). In the notation $\prod_{x,\nu}:= \prod_x \prod_{\nu=1}^D$ and $\sum_{n}:=\prod_{x,\nu} \sum_{n_{x,\nu}=0}^\infty$,
\begin{align}
e^{-S_{1}}=&\prod_{x,\nu} \exp{ g^{\nu\nu} \phi_{x+\nu} \phi_{x} }
=\sum_{n}\prod_{x,\nu} \frac{( g^{\nu\nu}\phi_{x+\nu} \phi_{x})^{n_{x,\nu}} }{n_{x,\nu}!} 
\\
=&\sum_{n}
(\prod_{x,\nu}\frac{( g^{\nu\nu})^{n_{x,\nu}}}{n_{x,\nu}! })
(\prod_{x} \phi_x^{\sum_{\nu=1}^D(n_{x,\nu}+n_{x-\nu,\nu})}),
\\
Z=\int D\phi ~ e^{-S}=& \mathcal{N} \sum_{n} (\prod_{x,\nu} \frac{( g^{\nu\nu})^{n_{x,\nu}}}{n_{x,\nu}!}) (\prod_{x} \int_{-\infty}^\infty d\phi_x ~\phi_x^{n_x} e^{-\eta \phi_x^2-V(\phi_x)}) \label{eq:rsfz21}
\\
=& \mathcal{N} \sum_{n} (\prod_{x,\nu} \frac{( g^{\nu\nu})^{n_{x,\nu}}}{n_{x,\nu}!}) (\prod_{x} f(n_x)),
\label{eq:rsfz22}
\end{align}
where $\mathcal{N}$ is a constant, $f$ stands for the last integral of \eqref{eq:rsfz21}, and $n_x:=\sum_{\nu=\pm 1}^{\pm D} n_{x,\nu}$. Here $Z$ is defined without the lattice spacing limit $a\rightarrow 0$ as for the toy model studied here I am contend with lattice results for simplicity.

When the theory exhibits a global $Z_2$ symmetry so that $V(r)=V(-r)$,
\begin{align}
f(n_x)=& \int_{-\infty}^\infty d\phi_x ~\phi_x^{n_x} e^{- \eta \phi_x^2- V(\phi_x)}
\\
=& \int_0^\infty d r_x ~ r_x^{n_x}e^{-\eta r_x^2}[e^{-V(r_x)}+(-1)^{n_x}e^{-V(-r_x)}]
\\
=& \int_0^\infty d r_x ~ r_x^{n_x}e^{-\eta r_x^2}[(1+(-1)^{n_x})e^{-V(r_x)}]
\\
=& 2 \delta_2(n_x)\int_0^\infty d r_x ~ r_x^{n_x}e^{-\eta r_x^2-V(r_x)},
\\
Z=& \mathcal{N}\sum_{n}(\prod_{x,\nu} \frac{(g^{\nu\nu})^{n_{x,\nu}}}{n_{x,\nu}!}) (\prod_{x} 2 \delta_2(n_x)\int_0^\infty d r_x ~ r_x^{n_x}e^{-\eta r_x^2-V(r_x)}).
\end{align}
The mod 2 Kronecker delta function $\delta_2(n_x)$ at all vertices implies that the particle lines keep extending and are never ending.

We can rewrite the path integral for a $Z_2$-symmetric theory in a tidier form
\begin{align}
Z=& \mathcal{N}\sum_{n\text{ extended}} \prod_{e} E_e(n_e) \prod_{v} V_v(n_v),
\end{align}
where $\sum_{n\text{ extended}}$ sums over extended particle configurations, $e$ relabels the lattice edges $(x,\nu)$, $v$ relabels the lattice vertices $x$, and
\begin{align}
E_e(n_e)=&\frac{(g^{e})^{n_{e}}}{n_{e}!},
\\
V_v(n_v)=&2\int_0^\infty d r ~ r^{n_v}e^{-\eta r^2-V(r)}
\end{align}
are the edge and vertex amplitudes. 
With the metric \eqref{eq:mrvpi}, $g^e=1$ identically.

\bibliographystyle{unsrt}
\bibliography{references.bib}

\end{document}